\documentclass[a4paper,twoside,10pt,fullpage]{article}
\newcommand{\affil}[1]{$^{\rm #1}$}
\usepackage[authoryear]{natbib}
\bibpunct{(}{)}{;}{n}{}{,}
\usepackage{graphicx}
\usepackage{rotating}

\date{} 

\title{\large\bf\flushleft Optical Corrections to the V\'eron-Cetty \& V\'eron Quasar Catalogue}
\author{\parbox{\textwidth}{\flushleft
\vspace{-0.5cm}
{\it E. Flesch\affil{A,B}}\\
\vspace{0.4cm}
{\small \affil{A}\,P.O. Box 12520, Wellington, New Zealand}\\
{\small \affil{B}\,Email: eric@flesch.org}}}

\begin{document}
\twocolumn[
\begin{changemargin}{.8cm}{.5cm}
\begin{minipage}{.9\textwidth}
\vspace{-1cm}
\maketitle
\small{\bf Abstract:  Fixes are presented to be applied to the V\'eron-Cetty \& V\'eron Quasar Catalogue, 13th edition.  These are comprised of 39 de-duplications, 380 astrometric moves of 8+ arcseconds of which 31 are over 10 arcminutes, and 30 indicated de-listings.}

\medskip{\bf Keywords:} catalogs --- quasars: general  

\medskip
\medskip
\end{minipage}
\end{changemargin}
]
\small

\section{Introduction}

Quasars have been catalogued into complete collections from the earliest days of quasar surveys; the two most prominent such catalogues have been the V\'eron-Cetty \& V\'eron (VCV) Catalogue of Quasars and Active Nuclei, published in 13 editions from 1984 to 2010, and the Hewitt \& Burbidge (HB) catalogue which published its last edition in 1993.  These catalogues recorded quasar positions based on optical or radio or X-ray surveys in which optical positions were not always published, or were often approximated into tiles of sky specified by the quasar names, e.g., `0450-132' referred to a tile of sky bounded by the corners B1950 04h50m-13d12m and 04h51m-13d18m (the last digit of the name being tenths of a degree).

With the advent of large optical quasar surveys like 2QZ (Croom et al. 2004) and SDSS(Abazajian et al. 2009), we have moved to an optical standard of quasar cataloguing where optical photometry and arcsecond-accurate astrometry are the norm.  To complete the transition to this new standard, there is accordingly a need to bring the older 1970s-1990s data into conformity.  Most of the old data already bears arcsecond-accurate astrometry, by which red-blue photometry can be sourced from optical catalogues like that of the Cambridge Automatic Plate Measuring machine (APM: McMahon \& Irwin 1992) and the United States Naval Observatory (USNO-B: Monet et al. 2003).  In their recent releases, VCV have fixed considerable old data up to the optical standard, but a residue remains of quasars with only approximately-known or mistaken positions.  Of course the latter of these are elusive to identify without some indicator to select them.

I have recently developed an unpublished but publically available quasar catalogue, the `Million Quasars' (Milliquas: Flesch 2012) catalogue, as a byproduct of the recently published `Atlas of Radio/X-ray Associations' (ARXA: Flesch 2010).  It is serving as a platform in support of ongoing development, and as a resource for bulk querying of quasars and quasar candidates.  It is built to be an optical catalogue of arcsecond accuracy, and I have relied on the VCV 13th edition as the authority on the early quasars.  Nevertheless, the process of merging the VCV data with optical APM/USNO-B data has highlighted some hundreds of quasars which either match to no optical signature, or to an untypical signature; this thus serves as the aforementioned indicator of quasar data which is poorly or mistakenly sited.  Investigation of these has led me to identify in the VCV data:

\begin{itemize}
	\item 39 duplicate objects where a correctly-sited quasar is matched by another, poorly-sited, object which arose either through error or inexactitude by the original author (OA) in the discovery paper, or as a result of a cataloguing error.
	\item 380 objects for which I find that the VCV position is offset at least 8 arcsec from the true optical position which I present.  31 of these quasars require a move of at least 10 arcminutes.
	\item 30 objects for which I recommend de-listing, as I have found, by a careful inspection of all evidence, either are not quasars, or have information of such poor quality that the object is lost beyond any mechanism of recovery, i.e., even if the object were re-found, it could not be recognized as being the previous object. 
\end{itemize}
  
These 449 objects represent all those where VCV objects were not well-supported by optical data, and found to require a fix.  For most of these objects, the diagnosis and correction have been clear.  But completeness requires me to include a few puzzling objects where the true identification is still unclear, for which I provide a `best' available optical object only, rather than something certain.  I flag such cases.  

\section{The Method}

The recent publications of the NBCKDE (Richards et al. 2009) and SDSS-XDQSO (Bovy et al. 2010) photometric quasar candidate catalogues, plus the radio/X-ray associated candidates published in ARXA, now highlight good optical candidates for quasar searches which hitherto would have been too hard.  These and other tools are:
\begin{itemize}
 	\item radio/X-ray associations to optical objects as presented in ARXA, and similarly in Milliquas which has updated versions of these. This is especially useful for originally X-ray selected quasars, such as from the {\it Einstein} satellite, which are thus expected to be X-ray associated in ARXA also.
	\item the photometric quasars published by NBCKDE and XDQSO (i.e. BOSS candidates).  NBCKDE provides reliable photometric redshifts, and some needed XDQSO redshifts were kindly provided by Adam Myers. 
	\item optical magnitudes and colours to match to the original paper, although the systematic offsets of that paper need to be ascertained; see the `personal equation' paragraph of section A5, Flesch \& Hardcastle 2004.
	\item recently surveyed quasars / galaxies of matching magnitude and redshift, especially those which are offset by right ascension only or declination only, as such errors do happen but rarely in both simultaneously.
	\item finding charts and positional information given by the original authors (OA), which later can have been overlooked or wrongly superseded.
	\item on-line optical look-ups, being the SDSS-DR7 \footnote{http://cas.sdss.org/dr7/en/tools/chart/chart.asp} and DR8 \footnote{ http://skyserver.sdss3.org/dr8/en/tools/chart/chart.asp} finding charts and the Digitized Sky Survey \footnote{DSS: http://archive.stsci.edu/cgi-bin/dss\_form}.
\end{itemize}

\begin{table*}
\scriptsize 
\caption{Objects found for Q1233+4749}
\begin{tabular}{rlllrrrll}
\hline 
   no. & J2000 & type & name & Rmag & Bmag & z & radio & X-ray \\
\hline
   1 & 12 35 22.9 +47 32 29 & QSO & Q 1233+4749 & & & 1.669 & & \\
   2 & 12 36 14.2 +47 32 59 & R/X & & 20.1 & 21.2 & & FIRSTJ123614.2+473258 & 1RXS J123614.8+473245 \\
\hline
\end{tabular}
\end{table*}

\begin{table*}
\scriptsize 
\caption{Objects found for Q1510+115}
\begin{tabular}{rlllrrrl}
\hline 
   no. & J2000 & type & name & Rmag & Bmag & z & radio \\
\hline
   1 & 15 12 23.8 +11 18 47 & QSO       & Q 1510+115               &      &      & 2.106 & \\
   2 & 15 12 37.2 +11 21 02 & photo-qso & NBCK J151237.20+112101.6 & 19.7 & 19.7 & 1.625 & \\
   3 & 15 12 49.2 +11 19 29 & QSO       & SDSS J151249.29+111929.3 & 17.4 & 17.5 & 2.110 & \\
   4 & 15 12 58.6 +11 18 49 & photo-qso & NBCK J151258.62+111849.0 & 19.6 & 20.5 & 2.225 & \\
   5 & 15 13 01.1 +11 23 23 & Galaxy    & SDSS J151301.10+112322.4 & 18.2 & 21.0 & 0.399 & FIRSTJ151301.0+112322 \\
   6 & 15 13 03.8 +11 20 49 & QSO       & SDSS J151303.83+112048.6 & 19.7 & 21.1 & 0.400 & \\
   7 & 15 13 15.0 +11 19 07 & photo-qso & NBCK J151314.97+111906.6 & 20.6 & 20.9 & 1.465 & \\
\hline
\end{tabular}
\end{table*}

\begin{table*}
\scriptsize 
\caption{NGC 157\#1: objects within 250 arcsec from MCG -1-02-034}
\begin{tabular}{rrlllrrr}
\hline 
   no. & dist & J2000 & name & type & z & Rmag & Bmag \\
\hline
  1 & 210  & 00 34 27.7 -07 55 18  & XDQ J003427.73-075517.5  & photo-qso &    & 22.4  & 22.3 \\
  2 & 119  & 00 34 33.9 -07 55 03  & BOSS J003433.95-075503.3  & photo-qso &    & 18.8  & 19.1 \\
  3 & 0  & 00 34 41.0 -07 54 09  & MCG -1-02-034  & Galaxy & 0.018  & 9.00  & 10.9 \\
\hline
\end{tabular}
\end{table*}

Sometimes these tools immediately yield a strong candidate. An example is my search for the optical object for Q 1233+4749, an approximately-sited quasar found serendipitously in the search for `primeval galaxies' by Thompson \& Djorgovski 1995, who presented no astrometry or photometry for this object, but their search was for objects of $19.5 \leq R < 24$ and $B-R \approx 1.5$.  This object should be within the tile of sky bounded by the corners B1233+4749 and B1234+4750; VCV placed it at the B1233+4749 corner.  Inspection of all available quasars and radio/x-ray associated objects and photometric quasar candidates within this tile of sky yielded only the two objects displayed in Table 1.  The first of these is the VCV-catalogued object which has no photometry and, being approximately-located, matches to no optical object.  The second object, offset 521 arcsec from the first, has R=20.1, B=21.2 at J123614.2+473259, i.e. B123351.6+474930, with both radio and X-ray associations -- the X-ray association is a {\it RASS} source with a 30-arcsec error circle.  ARXA assigns this object an 80\% likelihood of being a quasar based on the radio association and stellar psf.  This object has the right photometry for the discovery paper, and is clearly the true Q 1233+4749.  
 
Of course many searches for approximately-sited quasars are not so straightforward, and yield multiple candidates including some slightly outside the B1950 sky tile, but usually one candidate is found to be the clearly best match. A typical example is the search for Q 1510+155, from Sargent et al. 1988, which gives no photometry or astrometry, but states this is one of Cyril Hazard's quasars, which is typically a bluish $v = 17.5-18.5$.  Hazard had a large collection of unpublished quasars which he lent out to researchers, and this paper states that `the accurate positions of these objects will be reported (in) Sargent, Hazard, and McMahon 1988', which, however, never appeared.  The list of all quasars and candidates in the tile of sky bounded by B1510+155 and B1511+156 is displayed in Table 2.  There we first see the VCV object sited at the B1510+155 corner (although VCV usually positioned their approximate objects at the tile centres) with redshift of 2.106, then the 3rd object is an SDSS quasar of redshift 2.110 which is a very strong candidate for duplication.  Next is an NBCKDE candidate with photometric redshift of 2.225, also a good match, and the remaining objects are the wrong redshift.  To select between the SDSS quasar and the NBCKDE candidate we look at the object magnitudes: the SDSS quasar fits the Hazard profile of $v\approx 18$ and the NBCKDE candidate is two magnitudes too faint, so the SDSS quasar is selected as the true Q 1510+115.  This quasar is catalogued in VCV as `SDSS J15128+1119', so we have now found a duplication in VCV.  The Q 1510+115 row needs to be deleted, but I recommend to rename `SDSS J15128+1119' as `Q 1510+115' in order to retain the original name.

Sometimes there is no help from the radio/X-ray data, and no photometric survey objects.  For nine such approximately-sited cases there is only optical matching available, and for them I designate the best bluish object with magnitudes consistent to the discovery paper.  I flag these objects (in column 10 of Tables 5 - 7) according to how confident the selection is, and it is likely that I have missed the true object for some of these 9 cases.  

Some objects are corrected by identification of simple offsets, like IXO 15 which is at the true location J033311.3-361137 (Woo 2008), but recorded in VCV at J033311.9-371135, i.e., offset one degree to the south.  Such simple transcription errors account for $ \approx 50$ of the fixes presented here.

A few quasar discovery papers could be termed `treasure hunt' papers because the presented astrometry is not of quasars, but of radio/X-ray sources or galaxies with offsets supplied so that the reader must follow the trail to find the promised quasars.  Notable among these is Appenzeller et al. 1998, which lists 674 X-ray sources with offsets to objects which VCV diligently followed but ultimately missed 76 quasar positions which I herewith present.  Also notable is Arp 1981 which presents central galaxies with offsets provided to `companion' galaxies, and from there, further offsets to the described quasars.  Arp's treasure hunt was made do-able by his arcsecond-accurate offsets, and I present the search for his first object, `NGC 157\#1', as an example.  Arp's table 1 lists the secondary galaxy as offset 30 arcmin N of NGC 157 -- this will be MCG -1-02-034 which is 29.7 arcmin due north of NGC 157.  Arp now specifies a quasar of V=19.0 at a distance of 119 arcsec from the secondary galaxy, and I show, in Table 3, the list of photometric candidates within 250 arcsec of that galaxy.  There are 3 objects, the bottom of which is the galaxy, and the middle line shows a BOSS candidate at exactly 119 arcsec offset from the galaxy, R=18.8, B=19.1, which is undoubtedly Arp's quasar.  In this way I have identified seven object positions which VCV recorded only as `approximate'.  SDSS has since resurveyed these objects, finding one confirmed quasar, five photometric quasars, and one star.  

Many discovery papers present finding charts for their objects, which I've used in tandem with DSS to secure the exact astrometry.  The cutoff for inclusion in this paper is a move of 8 arcseconds.  Smaller moves are not always trivial, and are provided in the on-line Milliquas catalogue. 

\section{Duplicates}

39 duplicates in the VCV 13th edition are presented in Table 4.  Each duplication consists of a master object which has accurate optical astrometry, and the duplicate to be removed, which is typically positioned onto blank sky.  The offsets range from 48 arcsec to across the sky.  I find that 13 of these duplications arose because the original published astrometry was approximate only, 6 other cases were due to astrometric error by the original authors (OA), and 20 cases were cataloguing errors of which 11 were simple transcription errors causing a N/S or E/W offset.  Table 4 lists the duplicate first, then the master.  Columns are (1) line number, (2) VCV name of the duplicate object, (3) duplicate object J2000, (4) VCV redshift, (5) VCV V-magnitude, (6) VCV table number (1=QSOs, 2=Bl Lacs, 3=AGNe), (7) VCV reference number for the original paper, (8) astrometric offset from the duplicate to the master, (9) VCV name of the master object, (10) master object J2000, (11) master object redshift, (12) optical red magnitude, (13) optical blue magnitude, and (14) a comment explaining the original error and/or fix.  For 12 of these objects, I recommend to reassign the duplicate name to the master because the duplicate name is the original historic name.

\section{Astrometric moves}

380 VCV objects are found to be positioned 8+ arcseconds from their true optical positions.  I present these in four tables: Table 5 shows offsets of 30+ arcseconds, Table 6 shows offsets of 15-29 arcsec, Table 7 shows offsets of 8-14 arcsec, and Table 8 shows 82 objects which came late to hand, mostly from VCV table 3 of AGNe.  Columns are (1) line number, (2) VCV object name, (3) VCV J2000, (4) VCV redshift, (5) VCV V-magnitude, (6) VCV table number (1=QSOs, 2=Bl Lacs, 3=AGNe), (7) VCV reference number for the original paper, (8) astrometric move required, in arcseconds, (9) optical object J2000, (10) flag on optical object: p = a photometric quasar from NBCKDE/XDQSO, r = radio-associated, x = X-ray-associated, ! = a standout optical-only fit with no other good candidates, ? = best optical-only fit but other candidates present, ?? = good optical-only fit but other good candidates present, (11) optical red magnitude, (12) optical blue magnitude, and (13) a comment explaining the original error and/or fix. 

\section{Deletions}

I recommend to de-list 30 VCV-catalogued objects because they are either not quasars, or are so poorly described that recognition is excluded.  It's not suitable to make a table out of these objects because there are so many different circumstances to describe, so a simple list follows.  One paper accounts for 11 of these objects, which I'll discuss at the end.    

(1) NGC 3726 B1 is a star.  Approximately-located in VCV at J113448.1+470025, z=1.13, it is an Arp 1981 treasure hunt object, as discussed above.  Arp placed it at 100 arcsec from MCG 8-21-061, and it is seen exactly there, but SDSS-DR8 finds it to be a star, SDSS J113456.62+470014.8, r=17.9, b=18.5.  
             
(2) WEE 140 from Weedman 1985, putatively z=2.27, is found to be a star by SDSS-DR5, SDSS J160250.34 +280541.4.  Weedman did say in his Table 2 notes that the lines for this object were weak.

(3) Q 1052+04 from Lanzetta et al. 1991, \\ approximately-located in VCV at J105505.2+041400, v=18.1, z=3.391, does not exist.  NED \footnote{NASA/IPAC Extragalactic Database, http://ned.ipac.caltech.edu} identifies it with SDSS J105433.04+040027.4, z=3.301, but that object is 34 arcsec outside of the B1052+04 sky tile, and the redshift offset of 0.09 is large.  Another candidate is NBCK J105510.14+034730.0, v=18.5, which however has a photometric redshift of 0.535.  VCV's catalog paper explains (illustrated by its figure 2) that old z=3.3 redshifts were often wrong because low-z MgII-2800A lines were mistaken for Lyman-alpha.  In either scenario this object was wrongly presented.  No other good candidates are seen, and the original paper gives no help, so I recommend to de-list.   

(4) Q 0124-365 from Savage et al. 1984 is lost.  It was presented on an object-poor finding chart which I could not place anywhere in or near B0124-365, nor could VCV.  Perhaps there was a typo in the name, or maybe it was the wrong finding chart.  No astrometry was stated in the paper, so there is no means of recovering this object.

(5) MC 1227+120 never existed.  It was catalogued in Burbidge/Crowne/Smith 1977 with reference to a paper "in press", that paper being Smith et al. 1977, in which however it did not appear.  This object is only an artefact of the literature.

(6) 1ES 1249+174W is lost, if it ever existed.  It was presented as 1 of 2 optical candidates for a single {\it Einstein} source by Perlman et al. 1996.  The E object is an SDSS DR7 quasar, this W object, a "BL candidate" had no redshift presented -- the VCV-listed redshift belonged to the E object.  This W object was not measured and is unseen.
 
(7) 2E 1510+3902, an {\it Einstein} X-ray source at J151230.7+385051 with optical attributes of v=19.0, z=0.228, appears to be SDSS J151224.30+385112.7 of v=18.2, z=0.202; there are no other eligible objects.  The redshift is a poor match, and the discovery paper Reichert et al. 1982 which presented spectra for its new AGNe, stated problems with this object (Table 5, footnote b) and presented no spectrum.  Recommend de-listing due to poor fit and no other candidate.

(8) X404-23, from Zamorani et al. 1999, is an X-ray source without an optical object.  The finding chart presents a marked object which is, however, 30 arcsec outside the X-ray error circle.  As Zamorani et al. state, `None of them is a convincing identification'. 

(9) Q 0411-789, approximately located in VCV table 3 with V=16.0, z=0.019, is the only catalogued object from Campusano \& Pedreros 1978.  At such high latitude, the B1950 tile of sky is quite small, 3x1 arcmin, and no v=16 galaxy is seen there.  It is meant to be a radio-detected object, but the nearest SUMSS radio detection is 2 arcmin beyond the box. 

(10 \& 11) Two nameless table 3 entries, row 6502 (J081121.6+631943) and row 12235 (J101805.6+004318) are from Appenzeller et al. 1998, but they were identified as stars in that paper.
 
(12 \& 13) Two objects from Brissenden et al. 1987, approximately sited in VCV, are not seen at all.  They are 1H 0217-639 (J021906.4-634428, z=0.073) and 1H 2044-032 (J204711.6-030400, z=0.015).  These objects should be bright, $ V<18$, but no photometry is presented in a paper which does give photometry for its other objects.  1H 2044-032 was also presented with a 69 mJy radio detection, but the nearest NVSS source (5+ mJy) is 2 arcmin outside the box.

(14) IRAS 22040+0332 (J220634.2+034655, z=0.064, VCV table 3) from Gu et al. 1995, is not seen in or near the prescribed B1950 tile of sky.  No finding chart or optical magnitude was presented.  This IRAS source is not found in any IRAS source catalog.  No NVSS radio sources are within 5 arcmin.  

(15 \& 16) SDSS J09557+2525, z=2.262, and SDSS J10162+2649, z=0.383, were removed by SDSS-DR8 as they were satellite streaks, SDSS J095546.30+252534 and SDSS J101615.16+264902.4. 

(17) ROTSE J11568+5427, v=18.1, z=1.02, was presented on Astronomer's Telegram \#1515 \\ (http://www.astronomerstelegram.org) as an optical transient with suspected AGN origin.  However, subsequently Telegram \#1644 announced it to be ``an extremely luminous type II supernova at z=0.21''. 

(18 \& 19) Q 0000-029 (J000301.4-024141, v=18, z=2.31) and Q 2355+003 (J235803.4+004000, v=19, z=2.84) were so-called `optically violent variables' from Zhan \& Chen 1986, which appeared on a single UKST plate, but were not detected by deeper plates or CCD scans, and are today not seen on the DR8 finding charts.  The authors theorized these were AGNe but concluded "We cannot reject other explanations including galactic nova...".  Also to be considered are plate artefacts, since it is most unlikely to find two such optical transients on a single plate. 
        
(20-30) All quasars from Afanasiev et al. 1990, being SA68 \#110 (v=19.4, z=0.78), SA68 \#105 (v=20.6, z=0.71), SA68 \#090 (v=21.3, z=1.00), SA68 \#094 (v=19.3, z=1.08), SA68 \#143 (v=20.7, z=2.11), SA68 \#095 (v=21.3, z=1.24), M82 \#95 (v=19.4, z=1.01), M82 \#69 (v=19.4, z=0.93), M82 \#22 (v=19.6, z=0.96), SA57 \#216 (v=22.2, z=0.77), and SA57 \#431 (v=21.7, z=0.94).  Afanasiev presented these quasars from three fields, SA68 (i.e., `selected area 68') to a depth of $ B<22 $, M82 to $ B<22.5 $, and SA57 to $ B<23.5 $.  The stated astrometric accuracy is $ <2 $ arcsec.  All 3 fields are now covered by the SDSS DR8 finding charts, thus we are able to optically investigate Afanasiev's three fields to his plate limits.  Unfortunately, few if any of his quasars are seen.

Field SA68: Afanasiev presents 6 new quasars in this field, but five of them have no credible optical counterparts in the DR8 finding charts, while one of them, SA68 \#95, b=21.7, is offset 6 arcsec from SDSS J001735.30+155207.1, g=22.7, redshift unknown. This near-miss looks random.

Field M82: 7 quasars were presented, consisting of 4 already-known ones, HOAG 1, HOAG 2 \& HOAG 3 from Burbidge et al. 1980 and NGC 3031 U4 ({\it n\'ee} M82 \#4) from Arp 1983, and 3 new ones, M82 \#22, M82 \#69, and M82 \#95.  The 4 known quasars are all present in the DR8 finding charts, but the 3 Afanasiev objects are not seen at all.  

Field SA57:  7 quasars were presented, consisting of 5 already-known ones, KKC 30, KKC 36, KKC 37, KKC 41 \& KKC 43 from Koo/Kron/Cudworth 1986, and 2 new ones,  SA57 \#216 and SA57 \#431.  The 5 KKC quasars are all present in the DR8 finding charts, and here we finally see evidence of Afanasiev's quasars: SA57 \#216, b=22.6, matches exactly to \\ SDSS J130841.02+291857.4, g=22.5, redshift unknown.  In the case of SA57 \#431, Afanasiev mentions that this quasar is near another object, and indeed at that location there is a red object close to a blue object.  But the blue object is bright, v=17.3, while Afanasiev's quasar has v=21.7.  DSS POSS-I confirms the blue object was also v=17 in the 1950's epoch, so it is unreconcilably bright.

The final outcome is a head-scratcher, but the SDSS-DR8 finding charts are clear: there is only one credible match to Afanasiev's eleven objects.  Searches at increasing radii turn up nothing useful.  With this performance, the one apparent match should be discounted as a possible random artefact, therefore all of these should be de-listed.  Afanasiev published many other quasars in other papers, and those quasars are seen and confirmed and catalogued.

\begin{sidewaystable*}[p]
\addtolength{\tabcolsep}{-2pt}
\renewcommand{\arraystretch}{1.2}
\scriptsize
\caption{39 Duplications in VCV}
\begin{tabular}{r|llrrrr|r|llrrr|l}
\hline 
      &  name of  &       &   &  V  &     & OA  & offset & name of master & optical &   &  R  &  B  &         \\
   no & duplicate & J2000 & z & mag & tbl & ref & arcsec & optical object &  J2000  & z & mag & mag & comment \\
\hline
 1 & XMM J00005-2554   & 00 00 31.7 -25 54 59 & 0.283 &      & 2 & 1806 & 3602 & XMM J00005-2454 & 00 00 31.8 -24 54 57 & 0.284 & 15.6 & 17.4 & 1 deg N/S, original author (OA) error \\
 2 & TGS893Z369        & 00 23 20.1 -00 34 19 & 2.630 & 18.7 & 1 & 1421 & 144000 & 2GZ J002320-4034 & 00 23 20.0 -40 34 18 & 2.550 & 18.6 & 18.6 & 40 deg N/S offset, OA correct \\
 3 & IGR J00335+6126   & 00 33 34.8 +61 26 50 & 0.105 &      & 3 & 1745 & 124  & IGR J00333+6122  & 00 33 18.4 +61 27 43 & 0.105 & 18.7 & 20.6 & master is the publication of the dup \\
 4 & Q J0033-7546      & 00 33 49.3 -75 46 24 & 0.364 & 18.1 & 1 & 2311 & 19232 & MC4 0031-70 & 00 34 05.2 -70 25 52 & 0.363 & 17.1 & 16.9  & OA mispositioned to empty sky, \\ 
   &                   &                      &       &      &   &      &        &              &             &        &       &       & but stated dup to `MC40031-707' \\
 5 & BR 0035-25        & 00 37 58.7 -25 13 31 & 4.150 & 18.9 & 1 & 1106 & 8849 & BRI J0048-2442 & 00 48 34.5 -24 42 05 & 4.150 & 19.0 & 21.5 & OA typo, communication Mike Irwin \\
 6 & PSS J0052+2405    & 00 52 30.0 +24 05 30 & 1.900 & 17.4 & 1 & 1777 & 320 & PSS J0052+2405 & 00 52 06.6 +24 05 38 & 2.451 & 17.7 & 20.6 & approx pos dup of same name \\
 7 & SGP4:37           & 00 57 52.4 -27 45 17 & 1.691 & 21.0 & 1 & 283  & 794 & 2QZ J005852-2745 & 00 58 52.2 -27 45 18 & 1.689 & 20.6 & 20.6 & 1 time min offset, orig name SGP4:37 \\
 8 & SDSS J0112+0053   & 01 12 30.0 +00 53 29 & 4.566 & 20.2 & 1 & 1893 & 549147 & SDSSp J11228+0053 & 11 22 53.4 +00 53 31 & 4.560 &  & 23.4 & approx, VCV read J11228 as J011228 \\
 9 & TEX 0121+035      & 01 24 35.3 +03 47 31 & 0.221 & 19.6 & 3 & 90   & 1786 & MS 01200+0328 & 01 22 36.8 +03 43 54 & 0.221 & 17.9 & 19.0 & dup has co-ords of NGC 520, \\
   &                   &                      &       &      &   &      &        &              &             &        &       &       & FC Arp \& Duhalde 1985, `NGC 520.46'  \\
10 & 1H 0150-535       & 01 52 23.6 -53 20 34 & 1.560 & 17.0 & 1 & 306  & 1944 & H 0147-537 & 01 48 52.8 -53 28 29 & 1.568 & 17.1 & 17.9 & dup was approx located, dup identified by \\
   &                   &                      &       &      &   &      &        &              &             &        &       &       & Remillard et al. 1993, tbl 1, top row \\
11 & EIS J03325-2745   & 03 32 26.0 -27 45 30 & 1.221 & 21.3 & 1 & 538  & 53 & CDFS J03325-2745B & 03 32 30.0 -27 45 30 & 1.218 & 18.5 & 22.3 & 4 timesec offset, OA `KX4' correct \\
12 & CDFS 24           & 03 32 52.0 -27 52 04 & 3.590 &      & 1 & 819  & 135 & ECDF-S 475 & 03 32 41.8 -27 52 02 & 3.588 &  &  & 10 timesec offset, OA correct \\
13 & AX J0342.3-4412   & 03 42 19.4 -44 02 38 & 1.091 &      & 1 & 811  & 612 & XSF3:57 & 03 42 18.4 -44 12 50 & 1.091 &  & 21.5 & double transcription error, OA correct \\
14 & SDSS J03464+0037  & 03 46 29.0 +00 37 00 & 2.770 & 19.1 & 1 & 2380 & 802 & SDSS J03464+0023 & 03 46 29.0 +00 23 37 & 2.747 & 18.7 & 19.6 & decl truncation error  \\
15 & Q 0428-136        & 04 30 19.1 -13 29 32 & 3.244 & 20.8 & 1 & 616  & 476 & Q 0428-13        & 04 30 38.8 -13 35 52 & 3.249 & 17.2 & 18.7 & dup approx pos, master exactly loc \\
16 & Q 0450-132        & 04 52 18.9 -13 05 03 & 2.253 & 17.5 & 1 & 1996 & 800 & H 0450-1310 & 04 53 13.6 -13 05 55 & 2.250 & 17.2 & 17.7 & duplicate was approximately positioned, \\
   &                   &                      &       &      &   &      &        &              &             &        &       &       & new loc confirmed by Tytler et al. 2004  \\
17 & SDSS J07464+2449  & 07 46 25.8 +24 49 01 & 2.979 & 15.5 & 1 & 1957 & 3601 & SDSS J07464+2549 & 07 46 25.8 +25 49 02 & 2.979 & 19.2 & 19.9 & 1 deg N/S, transcription error \\
18 & Q 0837+109        & 08 39 43.6 +10 43 21 & 3.326 &      & 1 & 1996 & 734 & SDSS J08402+1034 & 08 40 17.8 +10 34 29 & 3.331 & 18.2 & 18.9 & dup approx pos, orig name Q 0837+109 \\
19 & LB 8814           & 08 53 14.5 +18 47 37 & 0.183 & 18.6 & 3 & 1016 & 115 & SDSS J08533+1847 & 08 53 22.4 +18 47 14 & 0.182 & 18.7 & 19.3 & dup approx pos, orig name LB 8814 \\
20 & NGC 2693 UB1      & 08 56 58.8 +51 20 45 & 2.310 & 19.5 & 1 & 84   & 192 & SDSS J08571+5123 & 08 57 11.1 +51 23 18 & 2.319 & 18.8 & 19.2 & Arp 1981 treasure hunt \& original name \\
21 & FIRST J09510+2210 & 09 12 03.8 +22 10 51 & 0.220 & 17.6 & 1 & 2528 & 33281 & SDSS J09520+2209 & 09 52 00.0 +22 09 06 & 0.220 & 17.4 & 17.3 & no such FBQS object, intended name\\
   &                   &                      &       &      &   &      &        &              &            &        &       &       &   was `FIRST J09520+2209', evidently \\
22 & SDSS J10479+0739  & 10 46 56.0 +07 39 52 & 0.168 & 19.3 & 3 & 1112 & 891 & SDSS J10479+0739 & 10 47 55.9 +07 39 52 & 0.168 & 19.5 & 20.5 & same object offset 1 time-min W \\
23 & RX J11062+0237    & 11 06 13.7 +02 37 58 & 0.564 & 17.1 & 1 & 2417 & 8001 & SDSS J11151+0237 & 11 15 07.6 +02 37 58 & 0.567 & 17.3 & 17.1 & E/W offset, OA correct at SDSS loc \\
24 & 2MASS J11178+2432 & 11 17 51.1 +24 32 08 & 0.088 & 17.9 & 3 & 1089 & 8187 & SDSS J11278+2432 & 11 27 51.1 +24 32 08 & 0.137 & 15.9 & 17.7 & 10 time-min E/W, OA loc correct \\
25 & PSS J1118+3702    & 11 18 56.2 +37 02 00 & 4.030 & 18.8 & 1 & 620  & 56 & SDSS J11189+3702 & 11 18 56.1 +37 02 56 & 4.025 & 19.2 & 21.4 & VCV omitted decl arcsecs, OA correct \\
26 & Q 1144+115        & 11 47 04.6 +11 16 20 & 2.438 &      & 1 & 2604 & 88 & SDSS J11469+1116 & 11 46 58.6 +11 16 19 & 2.512 & 18.3 & 19.0 & approx pos, Wolfe et al. 1986 say z=2.51 \\
27 & Q 1159+01         & 12 01 33.8 +00 43 18 & 3.269 &      & 1 & 1384 & 1980 & SDSS J12017+0116 & 12 01 44.4 +01 16 12 & 3.233 & 17.7 & 18.8 & dup approx pos, orig name Q 1159+01 \\
28 & Q 1211-1056       & 12 13 46.0 -11 13 14 & 1.626 & 19.0 & 1 & 1694 & 883 & TEX 1212-109 & 12 14 46.0 -11 13 13 & 1.626 & 19.3 & 19.7 & 1 time-min E/W, OA correct, `Q 1212-1056' \\
29 & KP 1229.8+07.7    & 12 32 22.1 +07 30 35 & 2.760 & 19.5 & 1 & 2179 & 48 & SDSS J12323+0731 & 12 32 19.7 +07 31 07 & 2.748 & 19.9 & 20.2 & OA finding chart \& original name \\
30 & Q 1333.3+2736     & 13 35 35.9 +27 27 18 & 0.783 & 19.0 & 1 & 515  & 354 & SDSS J13356+2721 & 13 35 36.0 +27 21 24 & 0.782 & 18.6 & 19.0 & 6 arcmin N/S, orig name Q 1333.3+2736 \\
31 & RXS J14194+4518   & 14 19 25.0 +45 18 32 & 0.076 & 16.7 & 3 & 2078 & 6330 & SDSS J14294+4518 & 14 29 25.0 +45 18 32 & 0.075 & 14.4 & 16.0 & 10 tmin E/W, transcription error  \\
32 & Q 1440-232        & 14 43 00.5 -23 29 15 & 2.221 & 18.0 & 1 & 355  & 6631 & PKS 1448-232 & 14 51 02.5 -23 29 31 & 2.221 & 17.2 & 17.5 & OA cited 1440-232 ``in preparation'', \\
   &                   &                      &       &      &   &      &        &              &             &        &       &       & but when published, it was 1448-232 \\
33 & RX J14462+2541    & 14 46 15.5 +25 41 43 & 0.188 & 17.3 & 1 & 2417 & 97107 & SDSS J16462+2541 & 16 46 15.5 +25 41 43 & 0.189 & 16.9 & 17.2 & 2 time-hours E/W, OA correct \\
34 & NGC 5866\#1       & 15 06 28.4 +55 45 30 & 0.706 & 18.1 & 1 & 79   & 449 & SDSS J15057+5549 & 15 05 43.9 +55 49 37 & 0.709 & 18.4 & 18.7 & dup approx pos, orig name NGC 5866\#1 \\
35 & Q 1510+115        & 15 12 23.8 +11 18 47 & 2.106 &      & 1 & 1996 & 377 & SDSS J15128+1119 & 15 12 49.2 +11 19 29 & 2.110 & 17.4 & 17.5 & dup approx pos, orig name Q 1510+115 \\
36 & Q 1511+091        & 15 13 25.9 +08 54 50 & 2.878 &      & 1 & 1996 & 400 & SDSS J15138+0855 & 15 13 52.5 +08 55 56 & 2.904 & 17.2 & 17.9 & dup approx pos, orig name Q 1511+091 \\
37 & SDSSp J1517+0101  & 15 17 30.0 +01 01 29 & 2.004 & 19.2 & 1 & 1893 & 335  & SDSS J15171+0101 & 15 17 07.6 +01 01 13 & 2.007 & 19.6 & 18.1 & dup approx pos, superseded \\
38 & TGS185Z206        & 23 00 37.7 +23 00 36 & 1.750 & 19.4 & 1 & 1421 & 179111 & 2GZ J230037-2644 & 23 00 37.7 -26 44 35 & 2.413 & 18.6 & 19.6 & Too far north for 2dF, VCV positioned \\
   &                   &                      &       &      &   &      &        &              &             &        &       &       & this object at RA-RA instead of RA-decl \\
39 & E 2352+073        & 23 55 20.5 +07 32 57 & 0.277 & 19.3 & 3 & 1872 & 5324 & RX J00013+0728 & 00 01 18.0 +07 28 28 & 0.270 & 18.3 & 18.5 & double transcription error by OA 
\footnote{This {\it Einstein} HRI-detected object was evidently misplaced and lost in preparation by the original authors, Reichert et al. 1982.  It was presented out-of-sequence in their RA-ordered table 4, as `2352+073' between `2353+283' and `2353+072', by which it is seen that the RA was originally inscribed as `2353'.  The presented {\it Einstein} position was (B1950) 23 52 47 +07 16 15 which is however only a 3-sigma detection without any eligible optical counterpart -- so the original optical object was not recovered.  The authors knew they had lost this object, for on their page 440, line 12, they wrote: "Four ... (HRI) counterparts" followed by only 3 names {\it sans} this object.  An X-ray associated {\it doppelganger} of the same redshift and similar magnitude, RX J00013+0728, is at (B1950) 23 58 44 +07 11 45, which would have been written as `2358+072'.  If that was the original object, then a double transcription error is indicated, causing the loss.} \\
\hline
\end{tabular}
\end{sidewaystable*}

\begin{table*}
\addtolength{\tabcolsep}{-2pt}
\renewcommand{\arraystretch}{1.0}
\tiny
\caption{105 moves of 30+ arcseconds}
\noindent\makebox[\textwidth]{%
\begin{tabular}{r|llrrrr|r|llrr|l}
\hline 
      &      &  VCV  &   &  V  &     & OA  & move & optical & opt &  R  &  B  &         \\
   no & name & J2000 & z & mag & tbl & ref & asec &  J2000  & typ & mag & mag & comment \\
\hline
  1 & PSS J0002+3732  & 00 02 46.1 +37 32 00  & 4.210  & 19.0  & 1  & 620  & 37  & 00 02 46.0 +37 32 37  & r  & 18.8  & 21.8  & VCV omitted decl arcsecs \\
  2 & PKS 0018-19  & 00 21 13.2 -19 10 44  & 0.095  & 17.0  & 3  & 1246  & 89  & 00 21 07.5 -19 10 06  & rx  & 13.6  & 16.0  & is galaxy PGC 1348, OA error, SIMBAD agrees \\
  3 & Q 0020-369  & 00 22 36.1 -36 40 22  & 2.005  & 19.2  & 1  & 88  & 54  & 00 22 37.9 -36 41 11  &    & 18.1  & 18.1  & FC of spectra \\
  4 & NGC 157\#1  & 00 34 46.3 -08 23 47  & 0.756  & 19.0  & 1  & 84  & 1734  & 00 34 33.9 -07 55 03  & p  & 18.8  & 19.1  & approx pos, Arp treasure hunt ok, BOSS cand \\
  5 & CS 47  & 00 49 30.2 -26 05 35  & 2.290  & 19.9  & 1  & 438  & 30  & 00 49 29.8 -26 05 05  &    & 19.1  & 20.1  & finding chart \\
  6 & Q 0047-2326  & 00 49 57.7 -23 09 40  & 3.422  &    & 1  & 1318  & 68  & 00 49 52.8 -23 09 49  & ! & 19.3  & 20.3  & approx pos, only good optical obj in B0047-2326 \\
  7 & Q 0112+029  & 01 14 34.9 +03 09 51  & 2.819  & 18.6  & 1  & 2341  & 410  & 01 14 53.2 +03 14 57  &    & 17.9  & 18.2  & approx pos, true pos courtesy of C. Ledoux, ESO \\
  8 & Q 0112-4556  & 01 14 49.2 -45 40 51  & 1.837  & 20.0  & 1  & 807  & 629  & 01 13 49.2 -45 40 51  &    & 19.6  & 20.3  & OA correct, VCV off by 1 time-min E \\
  9 & Q 0112-27  & 01 14 52.8 -27 14 09  & 2.896  &    & 1  & 1318  & 694  & 01 14 33.1 -27 24 52  & ?? & 18.2  & 19.0  & approx pos, bright Hazard, best opt only \\
 10 & Q 0112-381  & 01 15 08.4 -37 51 06  & 2.280  & 19.0  & 1  & 2015  & 150  & 01 15 08.2 -37 53 36  &    & 19.0  & 18.7  & hard-to-read FC (OA, plate 13), 150 asec S \\
 11 & Q 0128-367  & 01 30 41.8 -36 31 33  & 2.169  & 17.4  & 1  & 1261  & 33  & 01 30 42.2 -36 31 00  &    & 17.0  & 17.9  & move to 6dF J013042.3-363101, z=2.158 \\
 12 & Q 0129-369  & 01 32 13.4 -36 38 36  & 2.245  & 19.1  & 1  & 88  & 30  & 01 32 12.9 -36 39 05  &    & 18.3  & 18.3  & FC of spectra, standout in OA circled area \\
 13 & NGC 615 UB1  & 01 35 05.7 -07 20 29  & 1.640  & 18.5  & 1  & 84  & 249  & 01 34 52.1 -07 22 55  & prx  & 19.7  & 19.8  & approx pos, Arp treasure hunt ok, BOSS z ok \\
 14 & Q 0138-339  & 01 40 31.0 -33 40 51  & 2.356  & 18.5  & 1  & 88  & 58  & 01 40 30.2 -33 41 48  &    & 17.2  & 17.5  & FC of spectra \\
 15 & Q 0138-342  & 01 40 48.7 -33 57 51  & 2.074  & 18.3  & 1  & 88  & 68  & 01 40 50.2 -33 58 56  &    & 17.2  & 17.7  & FC of spectra \\
 16 & Q 0154-500  & 01 56 07.0 -49 45 31  & 2.460  & 18.7  & 1  & 2004  & 68  & 01 56 06.8 -49 44 23  &    & 19.1  & 19.3  & finding chart \\
 17 & NGC 772\#2  & 01 59 21.0 +19 00 32  & 2.612  & 19.4  & 1  & 101  & 338  & 01 59 39.2 +19 04 11  & p  & 19.3  & 19.6  & approx pos, opt pos redshift confirmed A. Myers \\
 18 & Q 0229+0656  & 02 32 04.5 +07 09 54  & 0.903  &    & 1  & 44  & 972  & 02 32 04.8 +07 26 06  & rx  & 17.8  & 18.0  & PKS 0229+072, OA given name was 0229+0712 \\
 19 & 2E 0237+3953  & 02 41 00.7 +40 07 21  & 0.528  & 18.3  & 1  & 2209  & 102  & 02 40 54.7 +40 06 06  & x  & 18.4  & 18.7  & {\it Einstein}, only X-ray object in vicinity \\
 20 & CFSM035  & 03 08 09.0 -55 09 07  & 0.590  &    & 1  & 1753  & 40  & 03 08 11.4 -55 09 41  &    & 18.4  & 19.1  & finding chart \\
 21 & IXO 15  & 03 33 11.9 -37 11 35  & 1.321  & 20.1  & 1  & 2613  & 3598  & 03 33 11.3 -36 11 37  & x  & 19.5  & 19.8  & FC, VCV wrote -37 instead of -36 \\
 22 & Q 0334-3554  & 03 36 37.2 -35 44 32  & 1.859  & 19.7  & 1  & 1566  & 2400  & 03 36 37.1 -35 04 32  & x  & 19.4  & 19.6  & OA correct, VCV off by 40 arcmin decl \\
 23 & Q 0336-3603  & 03 38 43.6 -35 53 49  & 0.741  & 19.6  & 1  & 1566  & 1200  & 03 38 43.5 -35 33 49  & x  & 19.3  & 19.3  & OA correct, VCV off by 20 arcmin decl \\
 24 & Q 0336-3507  & 03 38 53.0 -34 57 54  & 0.610  & 19.4  & 1  & 1566  & 1200  & 03 38 53.0 -34 37 54  & x  & 19.1  & 19.3  & OA correct, VCV off by 20 arcmin decl \\
 25 & Q 0337-3456  & 03 39 27.5 -34 47 07  & 1.364  & 19.7  & 1  & 1566  & 600  & 03 39 27.4 -34 37 07  & x  & 19.0  & 19.6  & OA correct, VCV off by 10 arcmin decl \\
 26 & XSF3:42  & 03 41 14.5 -44 05 27  & 2.277  & 21.9  & 1  & 2090  & 35  & 03 41 12.7 -44 05 56  & x  &    & 21.7  & 2RXP J034112.5-440601  \\
 27 & XSF3:20  & 03 41 38.8 -43 53 28  & 0.564  & 19.7  & 1  & 2090  & 621  & 03 42 36.2 -43 53 16  & x  & 18.9  & 19.2  & 1 tmin E, SIMBAD: 1AXG J034236-4352, \\
   &    &    &    &    &    &    &    &    &    &    &    & X-ray 2RXP J034235.6-435316 \\
 28 & RXS J03499+0640  & 03 49 59.7 +06 40 56  &    & 19.1  & 2  & 62  & 42  & 03 49 57.8 +06 41 27  & r  & 18.3  & 19.5  & Appenzeller treasure hunt ok \\
 29 & RXS J04173+0347  & 04 17 20.8 +03 47 01  & 0.082  & 17.3  & 3  & 62  & 34  & 04 17 18.6 +03 46 53  & r  & 16.8  & 17.6  & NVSS J041719.5+034650 \\
 30 & SDSSp J04375-0045  & 04 37 32.8 -00 45 18  & 2.818  & 22.8  & 1  & 936  & 150  & 04 37 42.8 -00 45 18  &    & 21.3  & 22.6  & 10 tsec E, OA ok, SDSS J043742.81-004517.6 \\
 31 & HE 0435-1223  & 04 38 14.8 -12 23 14  & 1.689  & 17.2  & 1  & 1622  & 359  & 04 38 14.8 -12 17 15  & x  & 16.4  & 16.2  & FC, VCV did not convert decl to J2000 \\
 32 & Q 0440-168  & 04 42 14.8 -16 42 22  & 2.679  & 18.0  & 1  & 1996  & 705  & 04 43 03.4 -16 43 55  & ? & 18.2  & 18.7  & approx pos, bright Hazard, best optical \\
 33 & RXS J06511+6842  & 06 51 06.8 +68 42 37  & 0.698  & 18.9  & 1  & 62  & 35  & 06 51 12.6 +68 42 51  & x  & 19.0  & 18.6  & Appenzeller treasure hunt ok \\
 34 & RXS J07181+6005  & 07 18 06.8 +60 05 33  & 1.018  & 20.4  & 1  & 62  & 38  & 07 18 09.8 +60 06 04  & x  & 19.7  & 20.8  & Appenzeller treasure hunt ok \\
 35 & RXS J07309+6343  & 07 30 50.6 +63 44 17  & 0.098  & 19.4  & 3  & 62  & 33  & 07 30 55.4 +63 44 09  & x  & 18.9  & 21.3  & 1RXS J073055.1+634405 \\
 36 & MCG +09.13.070  & 07 50 32.3 +56 29 02  & 0.019  & 18.0  & 3  & 1766  & 3600  & 07 50 28.9 +55 29 02  &    & 13.4  & 15.6  & 1 degree S \\
 37 & Q 0752+617  & 07 56 22.4 +61 34 01  & 1.892  &    & 1  & 20  & 65  & 07 56 31.4 +61 33 52  & ? & 17.5  & 18.3  & approx pos, best optical fit \\
 38 & RXS J07587+6334  & 07 58 44.8 +63 34 16  & 0.446  & 16.7  & 1  & 62  & 56  & 07 58 51.4 +63 33 41  & x  & 17.0  & 17.5  & Appenzeller treasure hunt ok \\
 39 & 1H 0828-706  & 08 28 17.2 -70 48 59  & 0.239  & 16.7  & 1  & 306  & 180  & 08 28 44.0 -70 51 02  & ?? & 16.3  & 17.1  & approx pos, best optical only \\
 40 & NGC 2639 C2.4  & 08 44 49.6 +54 23 06  & 2.630  &    & 1  & 352  & 14413  & 08 44 45.2 +50 22 53  & r  & 19.3  & 20.1  & OA correct, VCV off by 4 deg decl \\
 41 & WEE 16  & 08 45 39.9 +44 42 31  & 2.360  & 21.8  & 1  & 2488  & 56  & 08 45 35.6 +44 43 02  &    & 20.3  & 21.3  & finding chart \\
 42 & Q J08495+1150  & 08 49 35.4 +11 50 27  & 1.200  &    & 1  & 1750  & 35  & 08 49 36.6 +11 50 57  & p  & 19.0  & 19.7  & OA correct, VCV off by 30 asec decl \\
 43 & Q 0856+406  & 08 59 14.6 +40 28 19  & 2.290  &    & 1  & 2306  & 400  & 08 59 45.2 +40 25 05  & p  & 19.7  & 20.8  & approx pos, NBCKDE z ok, standout obj\\
 44 & NGC 2859 U3  & 09 24 54.2 +34 16 48  & 1.460  & 20.3  & 1  & 82  & 742  & 09 25 54.0 +34 16 45  & p  & 19.4  & 19.7  & FC, 1 time-min \\
 45 & NGC 2859 U2  & 09 24 57.8 +34 39 52  & 2.250  & 19.7  & 1  & 82  & 739  & 09 25 57.6 +34 39 50  & p  & 19.0  & 19.5  & FC, 1 time-min, but NBCKDE photo-z=0.735 \\
 46 & TB 0948+722  & 09 52 24.6 +71 57 55  & 0.529  &    & 1  & 18  & 138  & 09 52 54.2 +71 58 03  & x  & 17.2  & 17.2  & approx pos, standout X-ray obj \\
 47 & TB 0958+735  & 10 02 25.4 +73 15 32  & 2.067  & 17.0  & 1  & 17  & 227  & 10 03 17.6 +73 15 59  & rx  & 17.0  & 17.4  & approx pos, standout radio/X-ray obj \\
 48 & Q 1013+1126  & 10 16 09.7 +11 11 02  & 2.220  & 17.6  & 1  & 2238  & 416  & 10 16 37.9 +11 11 26  & ! & 17.6  & 18.2  & approx pos, standout optical match \\
 49 & SDSS J10292+2623A  & 10 29 15.9 +29 16 40  & 2.197  & 18.8  & 1  & 1100  & 10380  & 10 29 14.2 +26 23 40  & rx  & 18.4  & 18.6  & FC, lens component, `A' \& `B' switched \\
   &    &    &    &    &    &    &    &    &    &    &    & around, unfinished entry? \\
 50 & TOL 1036.7-27.2  & 10 38 49.0 -27 29 19  & 3.084  & 21.5  & 1  & 231  & 85  & 10 38 42.6 -27 29 12  &    & 18.7  & 20.6  & finding chart \\
 51 & NGC 3338 UB2  & 10 42 09.4 +13 45 53  & 2.140  & 19.7  & 1  & 84  & 1008  & 10 42 45.8 +13 31 37  & pr  & 19.9  & 19.9  & approx pos, Arp treasure hunt ok, NBCKDE z ok \\
 52 & NGC 3338 UB1  & 10 42 19.3 +13 44 53  & 2.040  & 20.4  & 1  & 84  & 1209  & 10 41 04.9 +13 53 48  & p  & 19.6  & 20.2  & approx pos, Arp treasure hunt ok, NBCKDE z ok \\
 53 & RXS J10478-0113  & 10 47 50.0 -01 12 43  & 0.435  & 18.9  & 1  & 62  & 41  & 10 47 51.6 -01 13 16  & x  & 16.9  & 18.6  & Appenzeller treasure hunt ok \\
 54 & RX J10509+5706  & 10 50 56.2 +57 06 48  & 0.550  & 20.5  & 3  & 2047  & 32  & 10 50 56.6 +57 07 20  & x  & 20.0  & 20.4  & OA object 67B `group' \\
 55 & RXS J10529-0736  & 10 52 59.6 -07 36 19  & 0.134  & 18.3  & 3  & 62  & 36  & 10 53 00.8 -07 35 48  &    & 17.3  & 18.2  & Appenzeller treasure hunt ok \\
 56 & NGC 3516 U2  & 11 03 57.4 +72 37 50  & 1.710  & 18.6  & 1  & 2238  & 58  & 11 03 45.4 +72 38 12  & x  & 18.3  & 19.2  & approx pos, standout object \\
 57 & Zw 011.012  & 11 09 42.9 -03 49 03  & 0.039  & 15.4  & 3  & 1246  & 576  & 11 09 41.5 -03 39 27  & x  & 12.0  & 13.8  & 10 arcmin S \\
 58 & LEDA 93604  & 11 26 02.4 +27 59 56  & 0.105  & 17.6  & 3  & 1169  & 17436  & 11 04 05.8 +28 00 15  &    & 15.6  & 16.8  & offset 22 time minutes from true location \\
 59 & IXO 40  & 11 50 57.9 -29 00 43  & 0.789  & 19.1  & 1  & 922  & 1002  & 11 50 57.8 -28 44 01  & x  & 20.0  & 19.6  & FC, 1000 arcsec N, transcription error \\
 60 & Q 1159+00  & 12 02 03.8 +00 13 18  & 2.580  &    & 1  & 1318  & 615  & 12 02 20.0 +00 22 42  &    & 17.2  & 17.6  & approx pos, move to Q120220.06+002242.1, \\
   &    &    &    &    &    &    &    &    &    &    &    & z=2.58, Crighton et al. 2011 \\
 61 & KP 1209.9+10.9  & 12 12 29.8 +10 41 23  & 2.100  & 21.0  & 1  & 2179  & 59  & 12 12 25.8 +10 41 30  &    & 20.2  & 20.5  & finding chart \\
 62 & KP 1227.8+07.4  & 12 30 22.7 +07 10 11  & 3.000  & 20.5  & 1  & 2179  & 68  & 12 30 27.0 +07 09 46  &    & 20.1  & 20.3  & finding chart \\
 63 & M 87.1524  & 12 31 02.2 +12 27 57  & 2.230  & 21.1  & 1  & 1635  & 468  & 12 30 30.8 +12 26 30  & px  &    & 19.7  & Strom treasure hunt {\it extraordinaire} ok, \\
   &    &    &    &    &    &    &    &    &    &    &    & Strom et al. 1981, NBCKDE photo-qso, X-ray \\
 64 & KP 1229.6+07.8  & 12 32 10.3 +07 33 44  & 1.510  & 20.0  & 1  & 2179  & 47  & 12 32 07.5 +07 34 05  & p  & 20.0  & 20.7  & finding chart \\
 65 & 2XMM J12321+2152  & 12 32 54.9 +21 52 55  & 1.870  &    & 1  & 580  & 696  & 12 32 04.9 +21 52 55  & x  &    &    & 50 time-sec offset, transcription error \\
 66 & Q 1233+4749  & 12 35 22.9 +47 32 29  & 1.669  &    & 1  & 2306  & 521  & 12 36 14.2 +47 32 59  & rx  & 20.1  & 21.2  & approx pos, standout obj, radio/X-ray, only \\
   &    &    &    &    &    &    &    &    &    &    &    & eligible object in narrow-decl tile of sky \\
 67 & 2MASS J13030-2447  & 13 03 02.0 -24 47 02  & 0.125  & 17.8  & 3  & 777  & 136  & 13 03 12.0 -24 47 02  &    & 18.0  & 19.6  & 10 timesec E to location given by OA \\
 68 & KP 1300.8+34.7  & 13 03 11.9 +34 27 09  & 1.900  & 21.0  & 1  & 2179  & 32  & 13 03 11.0 +34 26 39  & p  & 20.6  & 20.8  & finding chart \\
 69 & KP 1307.6+18.1  & 13 10 03.9 +17 53 42  & 1.900  & 21.0  & 1  & 2179  & 39  & 13 10 05.6 +17 53 11  & p  & 20.7  & 21.1  & finding chart \\
 70 & KP 1308.9+18.3  & 13 11 25.1 +18 08 04  & 1.710  & 20.0  & 1  & 2179  & 31  & 13 11 23.0 +18 08 13  & p  & 19.1  & 20.9  & finding chart \\
 71 & TOL 1313-309  & 13 16 28.7 -31 11 49  & 0.048  & 18.0  & 3  & 2146  & 57  & 13 16 32.6 -31 12 18  & r  & 14.6  & 16.6  & finding chart \\
 72 & VPM J13459+2741  & 13 45 44.5 +27 41 01  & 1.038  & 19.1  & 1  & 1563  & 135  & 13 45 54.6 +27 41 02  & p  & 18.9  & 18.9  & 10 tsec E \\
 73 & RX J13522-3228  & 13 52 16.0 -32 28 20  & 0.134  & 19.1  & 3  & 120  & 38  & 13 52 16.2 -32 28 58  & x  & 18.0  & 19.2  & 1RXS J135216.0-322850 is here, OA pos error \\
 74 & SMM J14010+0252  & 14 01 14.9 +02 52 23  & 2.565  &    & 1  & 2294  & 150  & 14 01 04.9 +02 52 24  &    & 21.0  & 22.0  & 10 tsec W \\
 75 & SBS 1401+566  & 14 02 42.7 +56 21 37  & 2.580  &    & 1  & 1480  & 107  & 14 02 55.4 +56 21 51  & ! & 17.0  & 18.1  & approx pos, best optical match in \\
   &    &    &    &    &    &    &    &    &    &    &    & B1401+566, OA says vmag 17.0-17.5 \\
 76 & 3C 295.0  & 14 11 20.5 +52 11 10  & 0.461  & 19.8  & 3  & 1579  & 60  & 14 11 20.4 +52 12 10  & rx  & 18.0  & 20.8  & 1 arcmin N, FIRST J141120.5+521209, CXO also \\
 77 & KP 1423.5+20.2  & 14 25 52.7 +20 02 43  & 2.190  & 20.5  & 1  & 2179  & 35  & 14 25 51.3 +20 03 12  & p  & 20.1  & 22.0  & finding chart \\
 78 & KP 1423.5+20.1  & 14 25 52.8 +19 56 34  & 1.840  & 20.5  & 1  & 2179  & 35  & 14 25 52.3 +19 57 08  & p  &    & 22.0  & finding chart \\
 79 & KP 1423.8+20.1  & 14 26 07.3 +19 57 48  & 2.300  & 20.5  & 1  & 2179  & 46  & 14 26 05.9 +19 58 30  &    & 20.3  & 21.1  & finding chart \\
 80 & NDWFS J14275+3522  & 14 27 39.7 +35 22 09  & 5.530  &    & 1  & 489  & 122  & 14 27 29.7 +35 22 09  &    &    &    & 10 timesec W to OA location \\
 81 & XBS J14496-0908  & 14 49 36.6 -09 08 29  & 1.260  & 19.3  & 1  & 372  & 65819  & 14 49 36.6 +09 08 30  & x  & 19.2  & 19.6  & into NH, change `-' to `+', OA pos error \\
 82 & KP 1504.0+21.7  & 15 06 16.2 +21 30 31  & 1.160  & 18.5  & 1  & 2179  & 30  & 15 06 18.0 +21 30 45  & p  & 20.2  & 20.6  & finding chart \\
 83 & Q 1532+2332  & 15 34 38.1 +23 22 30  & 1.249  & 19.8  & 1  & 106  & 137  & 15 34 48.0 +23 22 32  & px  & 20.0  & 21.1  & OA table 1, object `Arp 9', 10 tsec E \\
 84 & KISSR 854  & 15 45 26.8 +28 32 07  & 0.083  & 17.3  & 3  & 1127  & 3601  & 15 45 26.7 +29 32 08  &    & 15.6  & 17.3  & 1 degree N to AGN SDSS J154526.76+293207.7 \\
 85 & E 1550+721  & 15 49 40.6 +72 00 57  & 0.177  & 19.9  & 3  & 1872  & 36  & 15 49 45.6 +72 01 24  & x  & 17.6  & 18.7  & {\it Einstein} detected, this is the only X-ray object \\
   &    &    &    &    &    &    &    &    &    &    &    & in the vicinity, 2RXP J154945.3+720121 \\
 86 & RXS J16198+6926  & 16 19 51.8 +69 26 26  & 0.291  & 16.7  & 1  & 62  & 46  & 16 19 43.6 +69 26 41  & x  & 16.8  & 17.2  & Appenzeller treasure hunt ok \\
 87 & Q 1636+7403  & 16 35 16.2 +73 57 51  & 0.603  &    & 1  & 62  & 60  & 16 35 28.0 +73 57 17  &    & 18.2  & 18.5  & Appenzeller treasure hunt ok \\
 88 & RXS J16544+7007  & 16 54 24.0 +70 07 14  & 2.077  & 18.2  & 1  & 62  & 43  & 16 54 22.8 +70 06 31  & r  & 18.5  & 18.5  & Appenzeller treasure hunt ok \\
 89 & RXS J16596+7036  & 16 59 40.6 +70 36 32  & 0.835  & 18.9  & 1  & 62  & 39  & 16 59 39.6 +70 35 53  &    & 18.8  & 19.6  & Appenzeller treasure hunt ok \\
 90 & RXS J17196+7443  & 17 19 41.0 +74 43 00  & 1.500  & 19.1  & 2  & 62  & 36  & 17 19 39.8 +74 43 36  & pr  & 19.4  & 20.1  & Appenzeller treasure hunt ok \\
 91 & Q 1809+63  & 18 10 17.0 +63 29 14  & 1.091  &    & 1  & 825  & 59  & 18 10 17.0 +63 28 15  &    & 20.3  & 21.0  & OA error, wrote 63 29 14, is 63 28 14 ie 1 amin S \\
 92 & RX J18103+6328  & 18 10 23.6 +63 28 08  & 0.838  &    & 1  & 825  & 50  & 18 10 31.0 +63 28 09  & p  & 19.6  & 19.9  & OA section 6 gives opt pos, BOSS z ok \\
 93 & Pavo 1\#4  & 21 15 59.0 -67 57 15  & 0.408  & 19.8  & 3  & 895  & 360  & 21 15 58.4 -67 51 15  & x  & 18.1  & 19.7  & OA correct, VCV off by 6 arcmin decl \\
 94 & 2E 2141+0400  & 21 44 07.8 +04 13 25  & 0.401  & 20.6  & 3  & 2237  & 61  & 21 44 08.0 +04 14 26  & px  & 19.5  & 18.9  & FC, OA position restored, 1 arcmin N \\
 95 & MS 21432+1429  & 21 45 42.2 +14 42 55  & 1.387  & 20.4  & 1  & 1403  & 62  & 21 45 42.0 +14 43 57  & p  & 19.7  & 20.6  & FC, 1 arcmin S \\
 96 & Q 2217+0844  & 22 20 08.7 +09 00 02  & 0.228  & 17.6  & 1  & 955  & 60  & 22 20 08.6 +08 59 02  & r  & 17.6  & 18.4  & 1 arcmin S \\
 97 & Q 2239-386  & 22 42 21.7 -38 20 17  & 3.554  &    & 1  & 1318  & 183  & 22 42 37.0 -38 20 49  & x  & 19.5  & 20.7  & approx pos, Hazard hi-z, standout X-ray \\
 98 & B3 2311+396A  & 23 13 50.2 +40 03 03  &    & 18.5  & 2  & 2431  & 600  & 23 13 50.4 +39 53 03  &    & 19.1  & 20.2  & FC given but no position published, 10 amin S \\
 99 & NGC 7585 UB1  & 23 17 58.8 -04 39 54  & 1.410  & 18.7  & 1  & 84  & 558  & 23 18 16.6 -04 31 44  & px  & 18.1  & 18.5  & approx pos, Arp treasure hunt ok, BOSS z ok \\
100 & Q 2334+10  & 23 37 02.5 +10 46 36  & 2.243  &    & 1  & 1318  & 800  & 23 37 30.6 +10 58 00  & p  & 17.9  & 18.1  & approx pos, bright Hazard, BOSS z ok \\
101 & Q 2342+089  & 23 44 33.0 +09 10 39  & 2.784  &    & 1  & 1996  & 864  & 23 45 31.0 +09 09 06  & x  & 18.1  & 18.4  & approx pos, standout obj but 90 asec outside tile \\
102 & PSS J2344+3409  & 23 44 47.1 +34 09 00  & 3.960  & 19.0  & 1  & 620  & 41  & 23 44 47.0 +34 09 41  &    & 18.8  & 20.6  & VCV omitted decl arcsecs \\
103 & Q 2345-358  & 23 47 52.2 -35 32 01  & 2.386  & 17.9  & 1  & 88  & 48  & 23 47 51.1 -35 32 47  & x  & 17.8  & 18.0  & move to 6dF J234750.9-353248, z=2.387 \\
104 & Q 2351+10  & 23 54 03.4 +10 46 42  & 2.379  &    & 1  & 1318  & 833  & 23 53 41.0 +10 59 27  & p  & 18.4  & 18.4  & approx pos, bright Hazard, BOSS z ok, standout \\
105 & Q 2351-406  & 23 54 05.2 -40 23 38  & 2.100  & 21.5  & 1  & 1699  & 34  & 23 54 03.8 -40 24 08  &    & 19.7  & 20.4  & finding chart \\
\hline
\end{tabular}}
\end{table*}

\begin{table*}
\addtolength{\tabcolsep}{-2pt}
\renewcommand{\arraystretch}{1.0}
\tiny
\caption{89 moves of 15-29 arcseconds}
\noindent\makebox[\textwidth]{%
\begin{tabular}{r|llrrrr|r|llrr|l}
\hline 
      &      &  VCV  &   &  V  &     & OA  & move & optical & opt &  R  &  B  &         \\
   no & name & J2000 & z & mag & tbl & ref & asec &  J2000  & typ & mag & mag & comment \\
\hline
  1 & PC 0027+0520  & 00 30 30.6 +05 37 12  & 1.334  & 21.5  & 3  & 2061  & 15  & 00 30 29.6 +05 37 06  &    & 19.8  & 20.6  & finding chart \\
  2 & PC 0028+0453  & 00 31 10.1 +05 10 02  & 3.396  & 21.6  & 1  & 2061  & 15  & 00 31 09.6 +05 09 49  &    & 20.8  & 21.2  & finding chart \\
  3 & PC 00289+0451  & 00 31 12.1 +05 07 51  & 0.280  & 22.9  & 3  & 2061  & 15  & 00 31 11.6 +05 07 38  &    & 22.5  & 23.0  & finding chart \\
  4 & 1WGA J0057.3-2212  & 00 57 18.8 -22 12 34  &    & 20.6  & 2  & 1776  & 16  & 00 57 19.9 -22 12 38  & rx  & 21.0  & 21.6  & VCV gave X-ray position only \\
  5 & Q 0124-360  & 01 26 27.7 -35 45 19  & 1.560  &    & 1  & 2015  & 19  & 01 26 27.4 -35 45 00  &    & 17.8  & 18.3  & finding chart \\
  6 & Q 0127-371  & 01 29 40.8 -36 51 25  & 0.200  & 20.5  & 3  & 2015  & 16  & 01 29 40.6 -36 51 41  & r  & 18.8  & 19.1  & finding chart \\
  7 & Q 0149-3908  & 01 51 17.4 -38 53 48  & 2.250  & 19.5  & 1  & 1104  & 19  & 01 51 15.8 -38 53 41  & ! & 19.3 & 19.7 & standout optical match  \\
  8 & Q 02031+151  & 02 05 50.5 +15 23 25  & 2.000  & 20.3  & 1  & 45  & 16  & 02 05 50.4 +15 23 41  & p  & 19.7  & 20.7  & finding chart \\
  9 & FSM011  & 03 12 53.0 -55 09 28  & 0.658  &    & 1  & 1753  & 16  & 03 12 52.2 -55 09 42  &    &    & 21.4  & finding chart \\
 10 & XSF3:61  & 03 41 08.2 -44 15 26  & 1.730  & 21.9  & 1  & 2090  & 26  & 03 41 10.5 -44 15 32  & x  & 21.0  & 22.0  & 2RXP J034110.8-441533 \\
 11 & RXS J03411-3635  & 03 41 08.8 -36 34 59  & 0.347  &    & 3  & 2396  & 16  & 03 41 08.5 -36 35 15  & x  & 18.5  & 18.6  & corrected to OA position \\
 12 & XSF3:29  & 03 41 14.8 -43 57 27  & 0.660  & 22.3  & 3  & 2090  & 25  & 03 41 15.5 -43 57 03  & ?  &    & 22.3  & good optical, but near plate limit \\
 13 & XSF3:71  & 03 41 55.9 -44 22 29  & 1.080  & 21.5  & 1  & 2090  & 17  & 03 41 56.9 -44 22 16  & x  &    & 21.2  & 2RXP J034156.1-442214, SIMBAD agrees \\
 14 & XSF3:23  & 03 42 20.7 -43 55 30  & 2.551  & 22.1  & 1  & 2090  & 16  & 03 42 19.8 -43 55 17  & x  &    & 21.8  & 2RXP J034219.4-435519 \\
 15 & XSF3:56  & 03 43 55.9 -44 11 36  & 2.210  & 22.0  & 1  & 2090  & 18  & 03 43 56.1 -44 11 54  & ?  &    & 22.1  & best opt, 2RXP J034354.4-441203 nearby \\
 16 & H 0450-1310  & 04 53 12.8 -13 05 46  & 2.250  & 16.5  & 1  & 2344  & 15  & 04 53 13.6 -13 05 55  & x  & 17.2  & 17.7  & approx OA pos, standout X-ray \\
 17 & KP 0456.4+02.5  & 04 58 58.9 +02 35 44  & 1.430  & 17.5  & 1  & 2179  & 27  & 04 59 00.1 +02 35 24  &    & 18.2  & 18.9  & finding chart \\
 18 & RX J04598+1808  & 04 59 50.8 +18 08 38  & 0.157  & 19.5  & 3  & 1634  & 15  & 04 59 51.8 +18 08 40  & x  & 18.6  & 19.5  & only optical for 1RXS J045952.1+180831 \\
 19 & RXS J07161+5746  & 07 16 07.3 +57 46 03  & 0.488  & 19.6  & 3  & 62  & 20  & 07 16 07.2 +57 46 23  &    & 19.6  & 19.5  & Appenzeller treasure hunt ok \\
 20 & RXS J07177+6835  & 07 17 47.8 +68 35 37  & 1.115  & 17.7  & 1  & 62  & 19  & 07 17 46.8 +68 35 19  & x  & 17.7  & 17.9  & Appenzeller treasure hunt ok \\
 21 & RXS J07206+6338  & 07 20 39.0 +63 38 44  & 0.279  & 19.6  & 3  & 62  & 25  & 07 20 37.6 +63 39 07  &    & 19.3  & 20.3  & Appenzeller treasure hunt ok \\
 22 & RXS J07576+6557  & 07 57 36.4 +65 57 51  & 0.383  & 17.1  & 1  & 62  & 26  & 07 57 32.8 +65 58 04  &    & 20.1  & 21.0  & Appenzeller treasure hunt ok \\
 23 & 1WGA J0827.1+0841  & 08 27 06.0 +08 41 18  &    & 20.4  & 2  & 369  & 16  & 08 27 07.0 +08 41 21  & r  & 18.9  & 20.8  & standout radio Bl Lac cand \\
 24 & WEE 33  & 09 34 08.6 +21 29 08  & 2.310  & 21.1  & 1  & 2488  & 22  & 09 34 08.0 +21 29 28  & p  & 20.4  & 21.0  & finding chart \\
 25 & 1WGA J0953.9+4617  & 09 53 59.2 +46 17 18  & 0.612  & 19.8  & 1  & 1600  & 17  & 09 53 59.0 +46 17 01  & px  & 19.6  & 19.5  & standout photometric - X-ray \\
 26 & BOL 105  & 09 57 37.0 +69 38 12  & 2.240  & 21.4  & 1  & 237  & 19  & 09 57 40.2 +69 38 22  &    & 18.2  & 20.3  & finding chart \\
 27 & RXS J10304-0834  & 10 30 29.8 -08 34 05  & 0.207  & 16.6  & 1  & 62  & 22  & 10 30 30.9 -08 33 51  & x  & 18.1  & 18.7  & Appenzeller treasure hunt ok \\
 28 & TOL 1033.1-27.3  & 10 35 25.8 -27 33 42  & 1.610  & 21.8  & 1  & 231  & 19  & 10 35 25.6 -27 34 01  &    & 20.7  & 20.8  & finding chart \\
 29 & RXS J10384-0806  & 10 38 28.0 -08 06 13  & 0.216  & 15.7  & 1  & 62  & 22  & 10 38 28.4 -08 06 34  &    & 17.6  & 17.5  & Appenzeller treasure hunt ok \\
 30 & RXS J10478-0715  & 10 47 52.5 -07 15 10  & 0.095  & 18.3  & 3  & 62  & 24  & 10 47 51.4 -07 15 28  &    & 17.0  & 17.8  & Appenzeller treasure hunt ok \\
 31 & A4/22  & 11 02 05.8 +30 02 33  & 1.045  & 20.0  & 1  & 2027  & 22  & 11 02 07.2 +30 02 46  &    & 19.8  & 20.9  & only available optical obj, \\
    &    &    &    &    &    &    &    &    &    &    &    & OA table 2 stated 20 arcsec uncertainty \\
 32 & RXS J11028-0148  & 11 02 52.0 -01 48 51  &    & 21.7  & 2  & 62  & 23  & 11 02 53.0 -01 49 07  & prx  & 19.6  & 20.0  & Appenzeller treasure hunt ok \\
 33 & RXS J11054+0205  & 11 05 28.5 +02 05 19  & 0.247  &    & 3  & 62  & 16  & 11 05 29.5 +02 05 24  & x  & 18.2  & 19.6  & 2dF galaxy TGN435Z108, 1RXS J110528.6+020526 \\
 34 & WEE 43  & 11 16 07.6 +13 57 56  & 3.033  & 19.8  & 1  & 2488  & 21  & 11 16 06.4 +13 58 08  &    & 19.4  & 20.9  & finding chart \\
 35 & RXS J11167-1711  & 11 16 44.8 -17 11 27  & 0.375  & 17.5  & 1  & 120  & 24  & 11 16 43.5 -17 11 42  & ! & 16.9  & 17.7  & nearest optical, v ok \\
 36 & KP 1127.9+07.4  & 11 30 32.9 +07 12 32  & 2.000  & 20.5  & 1  & 2179  & 17  & 11 30 32.2 +07 12 46  &    & 18.7  & 19.1  & finding chart \\
 37 & KP 1128.7+07.4  & 11 31 18.5 +07 12 43  & 2.310  & 19.5  & 1  & 2179  & 28  & 11 31 16.8 +07 12 55  & p  & 19.6  & 20.1  & finding chart \\
 38 & 1E 1137+6555  & 11 40 16.6 +65 38 46  & 0.397  & 19.5  & 3  & 675  & 15  & 11 40 18.0 +65 38 34  & x  & 19.2  & 20.1  & finding chart \\
 39 & RXS J12124+2723  & 12 12 28.1 +27 23 43  & 0.376  & 18.9  & 1  & 62  & 19  & 12 12 28.7 +27 23 26  & x  & 18.7  & 19.3  & Appenzeller treasure hunt ok \\
 40 & 4C 26.37  & 12 24 33.0 +26 13 00  & 0.687  & 21.3  & 3  & 1776  & 15  & 12 24 33.2 +26 13 15  & pr  & 19.7 & 21.1 & standout photometric, radio lobes \\
 41 & KP 1229.0+07.8  & 12 31 35.8 +07 34 42  & 1.930  & 20.5  & 1  & 2179  & 27  & 12 31 34.0 +07 34 41  &    & 19.0  & 20.9  & finding chart \\
 42 & KP 1243.7+34.6  & 12 46 10.1 +34 21 02  & 2.290  & 20.0  & 1  & 2179  & 15  & 12 46 11.2 +34 21 08  & p  & 18.6  & 19.0  & finding chart \\
 43 & KP 1244.1+34.6  & 12 46 33.5 +34 24 42  & 1.900  & 20.0  & 1  & 2179  & 19  & 12 46 34.3 +34 24 58  & p  & 20.4  & 20.7  & finding chart \\
 44 & KP 1245.3+34.3  & 12 47 45.0 +34 02 06  & 1.820  & 20.0  & 1  & 2179  & 15  & 12 47 44.0 +34 02 15  & px  & 19.0  & 19.4  & finding chart \\
 45 & KP 1245.6+34.2  & 12 48 03.9 +34 00 13  & 2.070  & 20.5  & 1  & 2179  & 21  & 12 48 02.6 +34 00 26  & px  & 19.2  & 20.1  & finding chart \\
 46 & KP 1300.2+34.3  & 13 02 36.5 +34 04 57  & 1.800  & 20.5  & 1  & 2179  & 15  & 13 02 35.3 +34 04 56  &    & 19.6  & 21.6  & finding chart \\
 47 & KP 1300.5+34.6  & 13 02 53.4 +34 25 10  & 1.930  & 20.0  & 1  & 2179  & 19  & 13 02 53.2 +34 24 51  & p  & 20.6  & 20.6  & finding chart \\
 48 & AH 1  & 13 06 11.0 +29 30 02  & 0.710  & 19.0  & 1  & 798  & 15  & 13 06 11.0 +29 29 47  & ! & 18.9  & 19.4 & nearest optical, v ok \\
 49 & B8/19  & 13 09 58.7 +30 15 51  & 3.300  & 22.5  & 1  & 2027  & 19  & 13 09 58.0 +30 16 08  &    & 22.2  & 22.9  & only eligible object, u=25.7, \\
    &    &    &    &    &    &    &    &    &    &    &    & OA table 2 stated 20 arcsec uncertainty \\
 50 & WEE 120  & 13 11 43.0 +28 29 17  & 2.000  & 22.0  & 1  & 2488  & 26  & 13 11 43.2 +28 29 43  & p  & 20.2  & 20.8  & finding chart \\
 51 & PKS 1330-14  & 13 33 23.8 -14 36 21  &    & 19.0  & 2  & 1113  & 22  & 13 33 24.8 -14 36 06  &    & 19.1  & 20.7  & finding chart \\
 52 & IRAS 14054-1958  & 14 08 09.9 -20 12 15  & 0.161  & 22.0  & 3  & 2393  & 27  & 14 08 11.5 -20 12 29  & r  & 17.1  & 19.3  & OA gave right location and name \\
 53 & RX J14136-1538  & 14 13 40.3 -15 38 33  & 0.226  & 17.8  & 1  & 120  & 20  & 14 13 41.0 -15 38 15  & rx  & 17.1  & 18.0  & standout radio/X-ray object \\
 54 & 87GB 14208+2255  & 14 23 08.0 +22 42 19  & 4.316  & 18.7  & 1  & 191  & 21  & 14 23 08.2 +22 41 58  & prx  & 18.8  &    & standout photometric radio/X-ray \\
 55 & KP 1422.9+20.0  & 14 25 18.4 +19 48 20  & 2.000  & 20.0  & 1  & 2179  & 29  & 14 25 19.9 +19 48 40  & p  &    & 20.2  & finding chart \\
 56 & IRAS F14481+4454  & 14 49 56.3 +44 41 56  & 0.660  &    & 1  & 2572  & 29  & 14 49 53.6 +44 41 50  & pr  & 19.6  & 20.6  & standout photometric radio object, \\
    &    &    &    &    &    &    &    &    &    &    &    & OA says no X-ray \\
 57 & KP 1504.8+21.9  & 15 07 07.0 +21 48 07  & 2.140  & 20.5  & 1  & 2179  & 16  & 15 07 07.9 +21 47 58  & p  & 20.3  & 20.3  & finding chart \\
 58 & Q 15100-089  & 15 12 42.2 -09 09 01  & 2.100  & 20.4  & 1  & 45  & 29  & 15 12 42.4 -09 09 30  &    & 20.8  & 20.5  & finding chart \\
 59 & KP 1544.4+21.2  & 15 46 38.4 +21 02 58  & 2.000  & 20.5  & 1  & 2179  & 28  & 15 46 40.4 +21 02 54  & x  & 19.2  & 20.2  & finding chart \\
 60 & KP 1606.6+28.9  & 16 08 39.8 +28 51 46  & 2.560  & 19.0  & 1  & 2179  & 23  & 16 08 39.6 +28 51 23  &    & 19.6  & 19.9  & finding chart \\
 61 & KP 1607.3+28.8  & 16 09 21.4 +28 41 11  & 2.290  & 20.0  & 1  & 2179  & 28  & 16 09 19.8 +28 40 51  & p  & 19.3  & 20.0  & finding chart \\
 62 & KP 1622.0+26.8  & 16 24 08.4 +26 41 34  & 2.160  & 21.0  & 1  & 2179  & 29  & 16 24 10.4 +26 41 43  & px  & 20.0  & 20.9  & finding chart \\
 63 & RXS J16248+7039  & 16 24 48.4 +70 39 28  & 0.361  & 17.5  & 1  & 62  & 15  & 16 24 51.4 +70 39 25  & x  & 17.5  & 17.8  & Appenzeller treasure hunt ok \\
 64 & KP 1623.2+27.0  & 16 25 20.1 +26 53 34  & 2.300  & 20.0  & 1  & 2179  & 17  & 16 25 21.0 +26 53 22  & pr  & 20.2  & 20.8  & finding chart \\
 65 & RXS J16350+7009  & 16 35 02.2 +70 09 40  & 0.234  & 17.7  & 1  & 62  & 19  & 16 34 58.4 +70 09 40  & x  & 17.9  & 18.0  & Appenzeller treasure hunt ok \\
 66 & KP 1635.5+63.1  & 16 36 04.0 +62 59 06  & 1.920  & 20.0  & 1  & 2179  & 23  & 16 36 04.4 +62 58 43  & p  & 19.3  & 20.4  & finding chart \\
 67 & RXS J16362+7204  & 16 36 17.2 +72 04 57  & 0.187  & 16.8  & 1  & 62  & 18  & 16 36 15.2 +72 05 13  & x  & 16.2  & 17.8  & Appenzeller treasure hunt ok \\
 68 & KP 1635.5+26.6  & 16 37 37.2 +26 34 21  & 1.950  & 20.5  & 1  & 2179  & 27  & 16 37 35.2 +26 34 30  & px  & 19.9  & 20.8  & finding chart \\
 69 & RXS J16408+7448  & 16 40 49.4 +74 48 37  & 0.776  &    & 1  & 62  & 18  & 16 40 47.8 +74 48 20  &    & 18.7  & 17.8  & Appenzeller treasure hunt ok \\
 70 & RXS J17028+7222  & 17 02 51.2 +72 22 26  & 0.271  & 17.1  & 1  & 62  & 23  & 17 02 54.2 +72 22 45  & x  & 16.8  & 17.7  & Appenzeller treasure hunt ok \\
 71 & RXS J17070+7349  & 17 07 04.8 +73 49 31  & 0.916  & 18.2  & 1  & 62  & 18  & 17 07 01.0 +73 49 22  & pr  & 18.3  & 18.7  & Appenzeller treasure hunt ok \\
 72 & RXS J17190+7443  & 17 19 06.0 +74 42 59  & 1.530  & 17.9  & 1  & 62  & 16  & 17 19 05.8 +74 43 15  & p  & 18.6  & 18.2  & Appenzeller treasure hunt ok \\
 73 & Q 17257+503  & 17 26 57.6 +50 15 48  & 2.100  & 20.4  & 1  & 45  & 20  & 17 26 58.9 +50 16 04  & x  &    & 21.7  & finding chart \\
 74 & RX J18112+6543  & 18 11 09.6 +65 43 39  & 0.489  & 16.7  & 1  & 825  & 15  & 18 11 11.6 +65 43 48  & x  & 18.3  & 18.7  & galaxy NPM 1G+65.0166 is here, OA correct \\
 75 & KP 1826.8+48.5  & 18 28 06.9 +48 33 20  & 2.170  & 18.5  & 1  & 2179  & 27  & 18 28 08.2 +48 33 43  &    & 19.0  & 20.2  & finding chart \\
 76 & Q 1922+4748  & 19 23 26.0 +47 54 26  & 1.520  &    & 1  & 1609  & 16  & 19 23 27.2 +47 54 17  &    & 18.7  & 19.4  & MG4 J192325+4754 \\
 77 & RXS J20175-4114  & 20 17 31.2 -41 14 51  &    & 17.0  & 2  & 1246  & 28  & 20 17 30.0 -41 15 16  & r  & 18.6  & 20.6  & standout radio, VCV loc is X-ray loc, \\
    &    &    &    &    &    &    &    &    &    &    &    & {\it RASS} 1RXS J201731.2-411452 nearby  \\
 78 & 87GB 20451+2632  & 20 47 19.3 +26 43 40  & 1.280  & 20.6  & 1  & 736  & 27  & 20 47 20.3 +26 44 03  &    & 22.0  &    & lens, VCV did not report OA-given position \\
 79 & Q 2121+2459  & 21 23 34.4 +25 12 20  & 1.250  &    & 1  & 1609  & 19  & 21 23 34.7 +25 12 38  & rx  & 18.5  & 19.2  & standout radio/X-ray \\
 80 & Q 21265-150  & 21 29 18.9 -14 49 29  & 2.100  & 20.3  & 1  & 45  & 19  & 21 29 18.9 -14 49 48  &    & 19.4  & 20.0  & finding chart \\
 81 & Q 2142.8-1566  & 21 45 30.7 -15 25 41  & 2.050  & 21.2  & 1  & 2425  & 15  & 21 45 29.7 -15 25 44  & ! & 20.3  & 20.2  & best \& nearest optical \\
 82 & RXS J22103+0546  & 22 10 19.4 +05 46 58  & 0.150  &    & 3  & 62  & 18  & 22 10 18.6 +05 46 44  & r  & 16.2  & 18.3  & Appenzeller treasure hunt ok \\
 83 & RXS J22122+0351  & 22 12 16.2 +03 51 00  & 0.214  & 18.6  & 3  & 62  & 23  & 22 12 17.0 +03 50 41  & r  & 18.3  & 20.0  & Appenzeller treasure hunt ok \\
 84 & RXS J22179+0835  & 22 17 56.9 +08 35 43  & 0.206  & 19.2  & 3  & 62  & 16  & 22 17 55.8 +08 35 46  &    & 18.6  & 20.2  & Appenzeller treasure hunt ok \\
 85 & RXS J22211+1441  & 22 21 07.5 +14 41 04  & 0.894  & 18.3  & 1  & 62  & 19  & 22 21 08.0 +14 40 47  & px  & 18.2  & 18.1  & Appenzeller ok, BOSS z ok \\
 86 & RXS J22251+1136  & 22 25 11.2 +11 36 01  &    & 19.9  & 2  & 62  & 22  & 22 25 12.6 +11 36 02  & pr  & 19.6  & 20.3  & Appenzeller treasure hunt ok \\
 87 & Q 2226-3905  & 22 29 50.0 -38 50 00  & 1.133  & 18.6  & 1  & 857  & 19  & 22 29 51.4 -38 50 03  & ! & 17.6  & 18.5  & OA pos approx only, best optical \\
 88 & RXS J22321+0912  & 22 32 10.5 +09 12 18  & 0.383  & 17.5  & 1  & 62  & 15  & 22 32 11.4 +09 12 11  & x  & 17.7  & 17.7  & Appenzeller treasure hunt ok \\
 89 & RXS J22363+0756  & 22 36 20.7 +07 56 32  & 0.486  & 19.2  & 1  & 62  & 18  & 22 36 21.8 +07 56 38  & x  & 19.2  & 19.6  & Appenzeller treasure hunt ok \\
\hline
\end{tabular}}
\end{table*}

\begin{table*}
\addtolength{\tabcolsep}{-2pt}
\renewcommand{\arraystretch}{1.0}
\tiny
\caption{104 moves of 8-14 arcseconds}
\noindent\makebox[\textwidth]{%
\begin{tabular}{r|llrrrr|r|llrr|l}
\hline 
      &      &  VCV  &   &  V  &     & OA  & move & optical & opt &  R  &  B  &         \\
   no & name & J2000 & z & mag & tbl & ref & asec &  J2000  & typ & mag & mag & comment \\
\hline
  1 & Q 0004-019  & 00 07 22.5 -01 41 10  & 2.220  & 19.5  & 1  & 2673  & 8  & 00 07 22.0 -01 41 13  &    & 19.2  & 19.6  & finding chart \\
  2 & Q 0009+026B  & 00 12 32.3 +02 57 16  & 0.510  & 18.0  & 1  & 2672  & 10  & 00 12 32.4 +02 57 06  &    & 19.7  & 19.5  & finding chart \\
  3 & Q 0010+017  & 00 12 58.4 +02 03 07  & 2.240  & 18.0  & 1  & 2672  & 10  & 00 12 58.6 +02 02 57  &    & 18.8  & 19.3  & finding chart \\
  4 & TEX 0015+160  & 00 18 31.2 +16 20 44  & 1.333  &    & 1  & 2577  & 12  & 00 18 31.6 +16 20 55  & x  & 19.1  & 20.4  & finding chart \\
  5 & Q 0016-357  & 00 18 40.6 -35 29 13  & 3.190  & 19.1  & 1  & 1701  & 12  & 00 18 40.0 -35 29 04  & r  & 18.4  & 20.2  & standout radio \\
  6 & PC 0026+0501  & 00 28 40.0 +05 18 18  & 1.467  & 19.1  & 1  & 2061  & 8  & 00 28 40.3 +05 18 12  & x  & 18.6  & 19.0  & finding chart \\
  7 & PC 0028+0519  & 00 31 12.2 +05 35 47  & 0.080  & 19.3  & 3  & 2061  & 12  & 00 31 11.6 +05 35 39  &    & 19.4  & 20.1  & finding chart \\
  8 & PC 0028+0521  & 00 31 13.9 +05 38 08  & 0.331  & 21.9  & 3  & 2061  & 11  & 00 31 13.3 +05 38 01  &    & 21.8  & 22.5  & finding chart \\
  9 & Q J0035-7201  & 00 35 29.8 -72 01 23  & 0.666  & 19.4  & 1  & 2311  & 12  & 00 35 30.2 -72 01 35  & x  & 18.5  & 20.8  & finding chart \\
 10 & PKS 0041-395  & 00 43 37.6 -39 14 15  & 1.690  & 20.0  & 1  & 2003  & 9  & 00 43 36.8 -39 14 15  &    & 18.7  & 19.4  & finding chart \\
 11 & Q 0100-346  & 01 02 52.1 -34 25 17  & 0.800  &    & 1  & 2015  & 9  & 01 02 52.8 -34 25 14  &    & 18.1  & 18.4  & finding chart \\
 12 & Q 0126-330  & 01 29 15.5 -32 46 25  & 1.950  & 18.0  & 1  & 2015  & 9  & 01 29 14.9 -32 46 29  &    & 18.2  & 18.0  & finding chart \\
 13 & Q 0133-4122  & 01 35 23.2 -41 06 57  & 2.520  & 17.6  & 1  & 1104  & 12  & 01 35 22.1 -41 07 00  & ? & 17.2  & 19.9  & best stellar object \\
 14 & H 0143-0050  & 01 46 12.4 -00 35 39  & 3.100  & 17.0  & 1  & 1141  & 9  & 01 46 11.9 -00 35 33  &    & 16.4  & 18.1  & nearby bright \\
 15 & Q 0205-486U  & 02 07 47.5 -48 34 55  & 2.010  & 19.1  & 1  & 2004  & 9  & 02 07 47.6 -48 34 46  &    & 19.4  & 19.5  & finding chart \\
 16 & PC 0227+0018  & 02 30 11.3 +00 31 27  & 3.005  & 21.2  & 1  & 2061  & 9  & 02 30 11.0 +00 31 34  &    & 22.0  & 22.4  & finding chart \\
 17 & PC 0227+0029B  & 02 30 29.8 +00 42 57  & 2.576  & 20.7  & 1  & 2061  & 9  & 02 30 29.6 +00 43 05  & px  &    & 21.5  & finding chart \\
 18 & PC 0227+0053  & 02 30 31.6 +01 06 54  & 0.970  & 20.4  & 1  & 2061  & 12  & 02 30 31.2 +01 07 05  &    & 19.8  & 21.8  & finding chart \\
 19 & PC 0227+0024  & 02 30 31.7 +00 37 31  & 2.351  & 20.7  & 1  & 2061  & 8  & 02 30 31.6 +00 37 39  & px  & 19.6  & 20.2  & finding chart \\
 20 & PC 0227+0042  & 02 30 33.6 +00 56 15  & 3.892  & 20.7  & 1  & 2061  & 10  & 02 30 33.4 +00 56 25  &    & 21.4  & 23.0  & finding chart \\
 21 & PC 0228+0026  & 02 30 37.8 +00 39 40  & 1.346  & 19.8  & 1  & 2061  & 9  & 02 30 37.6 +00 39 48  & px  & 19.8  & 20.8  & finding chart \\
 22 & PC 0228+0114  & 02 30 41.2 +01 27 39  & 0.083  & 20.6  & 3  & 2061  & 14  & 02 30 40.9 +01 27 52  &    & 20.7  & 20.9  & finding chart \\
 23 & RXS J03313+0654  & 03 31 19.4 +06 54 28  &    & 19.2  & 2  & 62  & 11  & 03 31 19.1 +06 54 38  & px  & 18.8  & 19.3  & Appenzeller treasure hunt ok \\
 24 & RXS J04011+0004  & 04 01 10.8 +00 04 13  & 0.474  & 18.0  & 1  & 62  & 14  & 04 01 10.2 +00 04 02  &    & 18.2  & 18.7  & Appenzeller treasure hunt ok \\
 25 & MS 04200-3838  & 04 21 49.7 -38 31 26  & 0.831  & 21.1  & 3  & 1403  & 10  & 04 21 48.8 -38 31 25  & x  & 20.3  & 21.4  & finding chart \\
 26 & 2E 0438-1635  & 04 40 41.1 -16 29 51  & 0.520  & 19.8  & 3  & 675  & 11  & 04 40 41.8 -16 29 50  & x  & 18.9  & 20.5  & finding chart \\
 27 & H 0446-1440  & 04 48 36.2 -14 35 15  & 1.600  & 17.0  & 1  & 1141  & 8  & 04 48 36.7 -14 35 18  &    & 17.3  & 17.8  & standout v ok \\
 28 & 1ES 0446+449  & 04 50 07.3 +45 03 12  &    & 18.5  & 2  & 1775  & 10  & 04 50 06.6 +45 03 06  & x  & 13.5  & 20.3  & finding chart \\
 29 & H 0449-1325  & 04 51 42.6 -13 20 33  & 3.093  & 18.2  & 1  & 1997  & 14  & 04 51 42.7 -13 20 47  & ? & 17.6  & 18.9  & good v but OA position is approx, \\
    &    &    &    &    &    &    &    &    &    &    &    & alternative J045202.2-132049, r=17.7, b=19.3 \\
 30 & H 0449-1645  & 04 52 13.6 -16 40 12  & 2.600  & 17.0  & 1  & 1141  & 10  & 04 52 14.2 -16 40 16  &    & 17.4  & 17.1  & standout v ok \\
 31 & PKS 0524-433  & 05 25 54.5 -43 18 11  & 2.164  & 18.0  & 1  & 1790  & 12  & 05 25 55.0 -43 18 00  & r  & 18.0  & 19.1  & finding chart \\
 32 & 1E 0540+4935  & 05 43 56.0 +49 37 11  & 0.080  & 17.2  & 3  & 1008  & 10  & 05 43 56.8 +49 37 05  &    & 15.9  & 17.6  & finding chart \\
 33 & NGC 2403 S  & 07 34 57.2 +65 55 38  & 1.760  & 19.2  & 1  & 520  & 12  & 07 34 55.2 +65 55 39  & p  & 19.0  & 18.9  & finding chart \\
 34 & 1WGA J0814.3+0858  & 08 14 21.0 +08 57 00  & 0.240  & 19.9  & 2  & 369  & 12  & 08 14 21.5 +08 57 06  & r  & 19.3  & 20.0  & VCV pos is approx radio pos only, \\
    &    &    &    &    &    &    &    &    &    &    &    & Bl Lac behind galaxy, double lobes \\
 35 & AX J08362+5538  & 08 36 22.9 +55 38 53  & 1.290  & 21.4  & 1  & 1654  & 12  & 08 36 22.6 +55 38 41  & x  & 19.9  & 20.6  & X-ray obj, 2XMM J083622.6+553840 \\
 36 & WEE 20  & 08 46 19.9 +44 21 49  & 1.870  & 21.1  & 1  & 2488  & 8  & 08 46 20.2 +44 21 42  & p  & 19.2  & 19.3  & finding chart \\
 37 & KP 0847.2+15.5  & 08 50 02.4 +15 18 49  & 2.180  & 21.0  & 1  & 2179  & 8  & 08 50 02.1 +15 18 56  & p  & 19.9  & 20.7  & finding chart \\
 38 & KP 0847.6+15.6B  & 08 50 28.0 +15 28 22  & 2.200  & 19.0  & 1  & 2179  & 9  & 08 50 27.4 +15 28 21  & p  & 19.6  & 19.9  & finding chart \\
 39 & E 0907-091  & 09 09 36.2 -09 18 19  & 0.253  & 18.0  & 1  & 1447  & 9  & 09 09 36.1 -09 18 28  & x  & 18.1  & 18.4  & finding chart \\
 40 & IXO 32  & 09 10 19.9 +07 06 00  & 2.784  & 18.6  & 1  & 922  & 12  & 09 10 20.2 +07 05 49  & x  & 15.1  & 14.8  & finding chart \\
 41 & PC 0910+5625  & 09 14 37.9 +56 13 22  & 4.035  & 20.9  & 1  & 2045  & 12  & 09 14 39.3 +56 13 21  & x  & 20.8  & 22.9  & finding chart \\
 42 & RX J09553+4733  & 09 55 19.1 +47 34 25  & 1.730  & 20.0  & 1  & 2679  & 9  & 09 55 19.0 +47 34 16  & px  & 19.5  & 19.8  & finding chart \\
 43 & HOAG 2  & 09 57 19.8 +69 38 01  & 2.054  & 20.3  & 1  & 348  & 9  & 09 57 21.2 +69 37 55  & px  & 20.1  & 20.4  & finding chart \\
 44 & NGC 3031 U4  & 09 57 20.2 +69 35 37  & 0.850  & 20.1  & 1  & 86  & 12  & 09 57 22.4 +69 35 39  & x  &    & 20.6  & OA designated as `M82 \#4' \\
 45 & Q 1001-033  & 10 04 11.5 -03 36 42  & 0.458  & 19.5  & 3  & 1605  & 10  & 10 04 11.1 -03 36 50  & x  & 17.9  & 18.6  & finding chart \\
 46 & RXS J10336-1436  & 10 33 35.0 -14 36 24  & 0.367  & 20.2  & 2  & 369  & 12  & 10 33 35.8 -14 36 28  & rx  & 18.6  & 20.2  &  standout radio/X-ray \\
 47 & TOL 1038.2-27.1  & 10 40 33.5 -27 22 58  & 1.937  & 20.1  & 1  & 231  & 11  & 10 40 33.2 -27 23 08  &    & 18.1  & 19.9  & finding chart \\
 48 & RXS J10422-0715  & 10 42 14.8 -07 15 00  & 0.121  & 17.5  & 3  & 62  & 9  & 10 42 15.0 -07 15 09  & x  & 16.5  & 17.8  & Appenzeller treasure hunt ok \\
 49 & Q 1043+0539  & 10 45 46.7 +05 23 55  & 2.622  & 18.3  & 1  & 445  & 14  & 10 45 45.8 +05 23 56  &    & 18.3  & 19.5  & OA correct, VCV off by 1 tsec \\
 50 & RXS J10491-0647  & 10 49 07.2 -06 47 33  & 0.441  & 18.5  & 1  & 62  & 12  & 10 49 07.9 -06 47 38  &    & 18.2  & 18.2  & Appenzeller treasure hunt ok \\
 51 & RXS J10573-0805  & 10 57 18.6 -08 05 44  & 0.223  & 17.1  & 1  & 62  & 13  & 10 57 19.4 -08 05 40  & x  & 17.2  & 17.8  & Appenzeller treasure hunt ok \\
 52 & KP 1128.2+07.2  & 11 30 49.1 +07 00 49  & 2.120  & 19.0  & 1  & 2179  & 10  & 11 30 49.8 +07 00 47  & p  & 19.8  & 19.8  & finding chart \\
 53 & Q 1147+084  & 11 50 23.6 +08 08 56  & 2.627  &    & 1  & 2604  & 8  & 11 50 23.6 +08 08 48  & p  & 18.7  & 19.2  & photometric QSO  \\
 54 & KP 1209.2+10.7  & 12 11 45.4 +10 25 42  & 1.900  & 20.5  & 1  & 2179  & 9  & 12 11 45.0 +10 25 34  & p  & 20.3  & 20.9  & finding chart \\
 55 & MS 12152+2847  & 12 17 46.1 +28 31 19  & 2.830  & 20.0  & 1  & 1403  & 8  & 12 17 45.7 +28 31 25  &    & 20.1  & 20.6  & finding chart \\
 56 & Q 1217+085  & 12 20 04.8 +08 14 02  & 2.160  &    & 1  & 972  & 10  & 12 20 05.4 +08 14 02  & p  & 19.0  & 19.7  & NBCKDE photo-z = 2.165  \\
 57 & 1E 1227+024  & 12 29 52.8 +02 08 09  & 0.570  & 20.0  & 3  & 900  & 9  & 12 29 53.3 +02 08 12  & x  & 19.5  & 19.6  & matching X-ray object \\
 58 & 14A2  & 12 37 13.3 +01 23 00  & 0.722  & 19.4  & 1  & 282  & 9  & 12 37 13.2 +01 22 51  & p  & 18.7  & 18.8  & finding chart \\
 59 & KP 1244.0+34.5  & 12 46 30.1 +34 16 43  & 1.940  & 20.0  & 1  & 2179  & 8  & 12 46 30.4 +34 16 50  & p  & 18.9  & 19.3  & finding chart \\
 60 & KP 1258.4+34.2  & 13 00 52.4 +34 01 04  & 1.800  & 19.5  & 1  & 2179  & 12  & 13 00 52.8 +34 01 15  & p  & 20.4  & 20.6  & finding chart \\
 61 & KP 1259.4+34.7  & 13 01 46.5 +34 26 34  & 2.080  & 18.5  & 1  & 2179  & 12  & 13 01 47.1 +34 26 24  & p  & 19.2  & 19.5  & finding chart \\
 62 & 1RXS J13038-3950  & 13 03 50.6 -39 50 42  & 0.121  & 16.5  & 3  & 1814  & 9  & 13 03 50.5 -39 50 33  & rx  & 16.2  & 16.4  & standout radio/X-ray \\
 63 & TEX 1307+087  & 13 09 37.1 +08 28 19  &    & 17.9  & 2  & 1943  & 13  & 13 09 36.2 +08 28 15  & rx  & 19.4  & 20.8  & true radio/X-ray source, lobes, \\
    &               &                       &    &       &    &       &     &                       &     &  &    & OA evidently had J130941.1+082815, r=18.0, \\
    &               &                       &    &       &    &       &     &                       &     &  &    & which has X-ray but not radio, lobe nearby \\
 64 & IRAS 13155-0949  & 13 18 09.0 -10 05 08  & 0.104  &    & 3  & 1714  & 13  & 13 18 09.8 -10 05 12  & r  & 15.3  & 16.7  & NVSS J131809.7-100514 \\
 65 & Q 1339.1+2752  & 13 41 23.5 +27 37 02  & 0.463  & 20.0  & 3  & 516  & 11  & 13 41 24.1 +27 37 09  & x  & 19.7  & 20.0  & X-ray, v ok \\
 66 & Q 1339.9+2617  & 13 42 12.4 +26 01 56  & 2.504  & 20.1  & 1  & 516  & 9  & 13 42 12.2 +26 02 05  & p  & 19.8  & 20.8  & NBCKDE photo-z = 2.3 \\
 67 & Q 1402+2853  & 14 05 00.0 +28 39 04  & 1.430  & 20.1  & 1  & 956  & 11  & 14 05 00.1 +28 38 53  & px  & 18.7  & 19.6  & finding chart \\
 68 & Q 1407.3+2685  & 14 09 31.5 +26 36 38  & 1.900  & 20.9  & 1  & 511  & 11  & 14 09 31.5 +26 36 27  & p  & 19.9  & 21.0  & NBCKDE photo-z = 1.8 \\
 69 & Q J14172+2534  & 14 17 10.1 +25 34 39  & 0.852  & 18.4  & 1  & 347  & 9  & 14 17 10.0 +25 34 30  & px  & 18.7  & 19.4  & X-ray, NBCKDE photo-z = 0.855 \\
 70 & RXS J15002+1044  & 15 00 17.0 +10 44 41  & 0.114  & 17.7  & 3  & 2638  & 12  & 15 00 17.8 +10 44 41  & px  & 17.1  & 17.7  &   \\
 71 & KP 1528.9+14.4  & 15 31 16.8 +14 16 15  & 1.940  & 21.5  & 1  & 2179  & 13  & 15 31 16.2 +14 16 25  & p  & 20.3  & 20.1  & finding chart \\
 72 & MS 16011+4119  & 16 02 52.2 +41 11 50  & 0.534  & 19.7  & 1  & 1403  & 10  & 16 02 52.7 +41 11 58  & x  & 20.1  & 20.8  & finding chart \\
 73 & TEX 1615-040  & 16 18 37.4 -04 09 43  & 0.213  & 18.1  & 3  & 647  & 14  & 16 18 36.4 -04 09 43  & r  & 17.7  & 18.7  & FC, 1 timesec offset \\
 74 & RXS J16224+7038  & 16 22 28.4 +70 38 43  & 0.378  & 17.1  & 1  & 62  & 12  & 16 22 26.4 +70 38 36  & rx  & 17.1  & 17.6  & Appenzeller treasure hunt ok \\
 75 & KP 1623.4+26.9  & 16 25 28.3 +26 47 55  & 1.900  & 21.0  & 1  & 2179  & 12  & 16 25 28.8 +26 47 45  &    &    & 20.9  & finding chart \\
 76 & PC 1628+4948  & 16 29 26.6 +49 41 44  & 2.662  & 20.2  & 1  & 2042  & 11  & 16 29 27.5 +49 41 37  &    & 19.5  & 20.4  & finding chart \\
 77 & RXS J16299+7201  & 16 29 56.8 +72 01 24  & 0.609  & 18.2  & 1  & 62  & 9  & 16 29 54.8 +72 01 24  & rx  & 18.2  & 17.7  & Appenzeller treasure hunt ok \\
 78 & RXS J16315+7322  & 16 31 34.2 +73 22 22  & 1.001  & 18.9  & 1  & 62  & 11  & 16 31 36.4 +73 22 17  & x  & 19.0  & 19.4  & Appenzeller treasure hunt ok \\
 79 & RXS J16386+7040  & 16 38 39.2 +70 40 16  & 0.369  & 17.3  & 1  & 62  & 11  & 16 38 37.4 +70 40 23  & x  & 17.4  & 17.8  & Appenzeller treasure hunt ok \\
 80 & RXS J16427+7413  & 16 42 46.8 +74 13 48  & 0.664  & 19.1  & 1  & 62  & 8  & 16 42 47.2 +74 13 56  & x  & 19.2  & 19.5  & Appenzeller treasure hunt ok \\
 81 & RXS J16496+7442  & 16 49 41.4 +74 42 55  & 1.378  & 17.9  & 1  & 62  & 10  & 16 49 41.0 +74 42 45  & prx  & 18.4  & 18.4  & Appenzeller treasure hunt ok \\
 82 & RXS J16569+7110  & 16 56 55.4 +71 10 30  & 0.259  & 17.3  & 1  & 62  & 9  & 16 56 53.8 +71 10 34  & x  & 17.2  & 17.4  & Appenzeller treasure hunt ok \\
 83 & RXS J17125+7218  & 17 12 30.8 +72 18 33  & 0.241  & 19.6  & 3  & 62  & 8  & 17 12 29.8 +72 18 27  & x  & 18.3  & 20.5  & Appenzeller treasure hunt ok \\
 84 & PSS J1715+3809  & 17 15 39.6 +38 09 00  & 4.520  & 18.6  & 1  & 1837  & 10  & 17 15 39.4 +38 09 10  &    & 19.1  & 21.5  & VCV omitted decl arcsecs \\
 85 & Q 1715+535  & 17 16 35.2 +53 28 49  & 0.890  & 21.2  & 3  & 1659  & 13  & 17 16 36.6 +53 28 45  &    & 19.3  & 20.0  & finding chart \\
 86 & Q 1722.0+3474  & 17 23 47.9 +34 41 56  & 2.200  & 18.5  & 1  & 511  & 9  & 17 23 47.5 +34 41 48  & p  & 18.8  & 19.0  & only good v, but photo-z = 0.765 \\
 87 & RXS J17282+7253  & 17 28 17.6 +72 53 23  & 0.229  & 15.9  & 1  & 62  & 9  & 17 28 19.2 +72 53 28  & x  & 16.6  & 18.3  & Appenzeller treasure hunt ok \\
 88 & RXS J17349+7340  & 17 34 56.2 +73 40 03  & 1.131  & 17.8  & 1  & 62  & 10  & 17 34 54.6 +73 39 55  & x  & 17.9  & 18.0  & Appenzeller treasure hunt ok \\
 89 & 1WGA J1826.1-3650  & 18 26 08.1 -36 50 49  & 0.888  & 18.8  & 1  & 1312  & 9 & 18 26 08.3 -36 50 40  & x  & 16.3  & 17.8  & OA right, VCV off by 8 asec decl \\
 90 & Q 2111-4030  & 21 14 39.0 -40 17 43  & 1.750  & 20.9  & 1  & 1039  & 9  & 21 14 39.6 -40 17 48  &    & 20.4  & 20.6  & finding chart \\
 91 & 1E 21150+6027  & 21 16 19.0 +60 39 45  & 0.060  & 18.2  & 3  & 1008 & 20  & 21 16 16.4 +60 39 41  &   & 16.6  & 19.4  & finding chart  \\ 
 92 & Q 2130-0527  & 21 33 03.3 -05 14 02  & 0.329  &    & 3  & 1609  & 9  & 21 33 03.9 -05 14 03  & pr  & 17.2  & 18.0  & BOSS z ok \\
 93 & Q 2144.0-1589  & 21 46 43.4 -15 39 34  & 2.300  & 21.2  & 1  & 511  & 13  & 21 46 43.4 -15 39 21  & ! & 21.0  & 20.6  & standout nearest optical  \\
 94 & RXS J21549+0720  & 21 54 59.5 +07 20 02  & 0.352  &    & 3  & 62  & 14  & 21 55 00.0 +07 19 50  & rx  & 16.2  & 16.9  & Appenzeller treasure hunt ok \\
 95 & RXS J22204+0658  & 22 20 28.2 +06 58 11  & 0.116  & 17.7  & 3  & 62  & 11  & 22 20 27.6 +06 58 03  & x  & 17.3  & 18.5  & Appenzeller treasure hunt ok \\
 96 & MS 22225+2114  & 22 24 53.7 +21 30 01  & 0.617  & 17.2  & 1  & 1403  & 10  & 22 24 53.2 +21 30 09  & x  & 13.4  & 14.2  & finding chart \\
 97 & Q 2239.4+0045  & 22 42 00.0 +00 42 26  & 2.100  & 21.0  & 1  & 511  & 10  & 22 42 00.1 +00 42 36  & p  & 20.1  & 20.4  & NBCKDE photo-z = 2.185 \\
 98 & PB 7348  & 22 55 38.6 -11 14 53  & 1.330  & 17.5  & 1  & 1865  & 11  & 22 55 37.8 -11 14 52  &    & 16.8  & 17.6  & fits OA description of doublet \\
 99 & PC 2301+0021  & 23 04 14.6 +00 37 39  & 2.647  & 22.5  & 1  & 2042  & 9  & 23 04 15.1 +00 37 43  & p  &    & 22.2  & finding chart \\
100 & Q 2320-035W  & 23 23 26.4 -03 17 12  & 2.041  & 20.6  & 1  & 1790  & 9  & 23 23 27.0 -03 17 13  & p  & 19.2  & 19.5  & finding chart \\
101 & MS 23365+0517  & 23 39 07.0 +05 34 36  & 0.740  & 20.3  & 2  & 2149  & 11  & 23 39 07.4 +05 34 27  & prx  & 20.5  & 21.4  & finding chart \\
102 & BG CFH 56  & 23 49 54.7 +00 49 49  & 0.420  & 20.8  & 3  & 675  & 8  & 23 49 54.8 +00 49 57  &    & 20.5  & 20.9  & FC of 2347+005 in Ellingson/Yee/Green marks \\
   &    &    &    &    &    &    &    &    &    &    &    & wrong object (correct B1950 in their table 3) \\
103 & WEE 182  & 23 52 12.8 -01 20 21  & 2.362  & 20.3  & 1  & 2488  & 8  & 23 52 12.8 -01 20 29  &    & 18.0  & 18.0  & finding chart \\
104 & Q 2354-029  & 23 57 26.6 -02 42 08  & 2.170  & 19.6  & 1  & 2677  & 14  & 23 57 27.5 -02 42 09  &    & 19.2  & 18.7  & finding chart \\
\hline
\end{tabular}}
\end{table*}

\begin{table*}
\addtolength{\tabcolsep}{-2pt}
\renewcommand{\arraystretch}{1.0}
\tiny
\caption{82 additional objects}
\noindent\makebox[\textwidth]{%
\begin{tabular}{r|llrrrr|r|llrr|l}
\hline 
      &      &  VCV  &   &  V  &     & OA  & move & optical & opt &  R  &  B  &         \\
   no & name & J2000 & z & mag & tbl & ref & asec &  J2000  & typ & mag & mag & comment \\
\hline
  1 & PC 0026+0522     & 00 28 58.8 +05 39 09  & 0.101  & 20.7  & 3  & 2061 & 8  & 00 28 59.2 +05 39 06  &    & 19.6  & 20.9  & finding chart \\
  2 & PC 0026+0457     & 00 29 02.7 +05 14 18  & 0.300  & 20.5  & 3  & 2061 & 9  & 00 29 03.2 +05 14 12  &    & 19.8  & 20.9  & finding chart \\
  3 & PC 0026+0455     & 00 29 12.7 +05 11 54  & 0.435  & 21.5  & 3  & 2061 & 8  & 00 29 13.1 +05 11 49  &    & 22.3  & 23.0  & finding chart \\
  4 & PC 0027+0532     & 00 29 44.7 +05 49 13  & 0.069  & 21.2  & 3  & 2061 & 11  & 00 29 44.6 +05 49 24  &    & 20.1  & 21.9  & finding chart \\
  5 & PC 0027+0459A    & 00 29 59.2 +05 15 47  & 0.153  & 19.9  & 3  & 2061 & 8  & 00 29 59.2 +05 15 55  &    & 19.5  & 19.7  & finding chart \\
  6 & PC 0027+0456     & 00 30 20.8 +05 12 52  & 0.079  & 21.2  & 3  & 2061 & 17  & 00 30 19.8 +05 12 43  &    & 19.6  & 21.1  & finding chart \\
  7 & PC 0028+0529     & 00 30 40.1 +05 46 03  & 0.327  & 21.0  & 3  & 2061 & 13  & 00 30 39.4 +05 45 56  &    & 19.6  & 20.9  & finding chart \\
  8 & PC 0028+0522     & 00 31 02.8 +05 38 52  & 0.046  & 18.5  & 3  & 2061 & 12  & 00 31 02.0 +05 38 47  &    & 17.8  & 18.3  & finding chart \\
  9 & PC 0028+0505     & 00 31 07.2 +05 22 21  & 0.430  & 22.7  & 3  & 2061 & 14  & 00 31 06.6 +05 22 10  &    & 22.7  & 23.3  & finding chart \\
 10 & PC 0028+0512     & 00 31 10.5 +05 28 46  & 0.171  & 22.4  & 3  & 2061 & 13  & 00 31 09.9 +05 28 36  &    &    & 21.7  & finding chart \\
 11 & PC 0028+0519     & 00 31 12.2 +05 35 47  & 0.080  & 19.3  & 3  & 2061 & 12  & 00 31 11.6 +05 35 39  &    & 19.4  & 20.1  & finding chart \\
 12 & PC 0028+0455     & 00 31 14.7 +05 12 10  & 0.228  & 21.2  & 3  & 2061 & 14  & 00 31 14.2 +05 11 58  &    & 20.4  & 21.9  & finding chart \\
 13 & MS 00387+3251    & 00 41 26.2 +33 07 49  & 0.225  & 18.5  & 3  & 823  & 17  & 00 41 25.5 +33 08 04  & x  & 17.3  & 18.4  & finding chart \\
 14 & 1WGA J0057.6+3022 & 00 57 38.5 +30 22 39 & 0.180  & 21.6  & 3  & 1230 & 8  & 00 57 38.6 +30 22 47  & x  & 19.8  & 20.7  & standout X-ray \\
 15 & TGS132Z079       & 01 53 34.3 -25 52 16  & 1.070  & 20.5  & 1  & 1421 & 96961 & 23 53 34.4 -25 52 14  &    & 18.6  & 19.8  & VCV off by 2 time hours, to 2dF position \\
 16 & PC 0227+0056     & 02 30 09.5 +01 09 50  & 0.186  & 20.1  & 3  & 2061 & 13  & 02 30 09.2 +01 10 02  &    & 19.4  & 20.2  & finding chart \\
 17 & PC 0227+0100A    & 02 30 10.5 +01 14 13  & 0.167  & 22.2  & 3  & 2061 & 13  & 02 30 10.0 +01 14 24  &    &    & 21.9  & finding chart \\
 18 & PC 0227+0104     & 02 30 14.7 +01 17 27  & 0.074  & 20.4  & 3  & 2061 & 14  & 02 30 14.3 +01 17 40  &    & 20.3  & 20.7  & finding chart \\
 19 & PC 0227+0012     & 02 30 14.9 +00 26 05  & 0.071  & 19.2  & 3  & 2061 & 8  & 02 30 14.6 +00 26 12  &    & 18.5  & 19.8  & finding chart \\
 20 & PC 0227+0011     & 02 30 16.5 +00 24 34  & 0.149  & 19.4  & 3  & 2061 & 8  & 02 30 16.2 +00 24 41  &    & 19.2  & 20.0  & finding chart \\
 21 & PC 0227+0034     & 02 30 18.5 +00 47 52  & 0.025  & 20.9  & 3  & 2061 & 10  & 02 30 18.2 +00 48 01  &    & 20.6  & 21.1  & finding chart \\
 22 & PC 0227+0029A    & 02 30 24.4 +00 42 43  & 0.081  & 17.9  & 3  & 2061 & 11  & 02 30 24.0 +00 42 53  &    & 16.8  & 17.3  & finding chart \\
 23 & PC 0227+0109     & 02 30 27.5 +01 23 13  & 0.178  & 20.4  & 3  & 2061 & 14  & 02 30 27.2 +01 23 26  &    & 19.8  & 20.7  & finding chart \\
 24 & RXS J02325+3404  & 02 32 33.1 +34 04 28  & 0.079  & 16.3  & 3  & 170  & 11  & 02 32 32.3 +34 04 23  & x  & 15.6  & 16.0  & finding chart \\
 25 & MS 03403+0446    & 03 43 04.8 +04 55 45  & 0.190  & 20.0  & 3  & 1403 & 10  & 03 43 05.4 +04 55 45  &    & 19.0  & 20.1  & finding chart \\
 26 & RXS J03535+0204  & 03 53 34.1 +02 04 37  & 0.247  & 18.4  & 3  & 62   & 21  & 03 53 34.4 +02 04 57  & x  & 18.1  & 18.5  & Appenzeller treasure hunt ok \\
 27 & RXS J04027+0159  & 04 02 42.0 +01 59 54  & 0.151  & 17.7  & 3  & 62   & 13  & 04 02 42.2 +02 00 07  & x  & 17.2  & 18.5  & Appenzeller treasure hunt ok \\
 28 & B2 0402+37       & 04 05 49.3 +38 03 32  & 0.054  & 18.5  & 3  & 2560 & 15  & 04 05 48.1 +38 03 26  & x  & 17.4  & 20.0  & finding chart \\
 29 & RXS J04173+0347  & 04 17 20.8 +03 47 01  & 0.082  & 17.3  & 3  & 62   & 17  & 04 17 19.8 +03 46 53  & r  & 16.8  & 17.6  & Appenzeller treasure hunt ok \\
 30 & IRAS 04413+2608  & 04 44 28.6 +26 14 04  & 0.171  &       & 3  & 362  & 31  & 04 44 30.8 +26 14 10  & r  & 17.7  & 19.9  & finding chart \\
 31 & IRAS 04493-6441  & 04 49 40.8 -64 36 11  & 0.060  & 15.4  & 3  & 575  & 10  & 04 49 40.8 -64 36 21  &    & 15.3  & 14.7  & finding chart \\
 32 & RC J0457+0452    & 04 57 54.3 +04 53 48  & 0.186  & 19.2  & 3  & 628  & 9  & 04 57 53.8 +04 53 54  & r  & 18.5  & 21.2  & standout radio, SIMBAD agrees \\
 33 & RXS J07157+5912  & 07 15 46.3 +59 12 33  & 0.288  & 19.1  & 3  & 62   & 31  & 07 15 44.2 +59 12 59  &    & 17.5  & 19.9  & Appenzeller treasure hunt ok \\
 34 & RXS J07207+6543  & 07 20 44.2 +65 43 47  & 0.483  & 19.7  & 3  & 62   & 35  & 07 20 49.0 +65 44 05  & r  & 19.5  & 20.1  & Appenzeller treasure hunt ok \\
 35 & RXS J07281+6718  & 07 28 11.4 +67 18 14  & 0.135  & 18.0  & 3  & 62   & 8  & 07 28 12.8 +67 18 15  & x  & 17.1  & 18.8  & Appenzeller treasure hunt ok \\
 36 & IRAS 07246+6125  & 07 29 09.4 +61 18 53  & 0.137  & 19.2  & 3  & 1327 & 20  & 07 29 12.2 +61 18 53  & r  & 17.7  & 18.9  & 3 tsec E, 2008 ApJ 683, 114 \\
 37 & RXS J08111+5730  & 08 11 09.1 +57 30 03  & 0.082  & 19.3  & 3  & 62   & 22  & 08 11 08.3 +57 30 24  & rx  & 18.3  & 20.1  & Appenzeller treasure hunt ok \\
 38 & RXS J08162+6600  & 08 16 17.8 +66 00 52  & 0.251  & 19.0  & 3  & 62   & 21  & 08 16 21.2 +66 00 50  & rx  & 18.3  & 20.1  & Appenzeller treasure hunt ok \\
 39 & RXS J08244+6249  & 08 24 26.8 +62 49 27  & 0.097  & 18.1  & 3  & 62   & 13  & 08 24 27.8 +62 49 38  & rx  & 18.4  & 19.5  & Appenzeller treasure hunt ok \\
 40 & Zw 064.024       & 09 59 46.8 +11 28 20  & 0.077  & 15.7  & 3  & 1246 & 59  & 09 59 46.8 +11 29 19  & x  & 10.6  & 12.4  & 1 arcmin S \\
 41 & RXS J10197-0016  & 10 19 44.3 -00 16 29  & 0.075  & 19.4  & 3  & 62   & 11  & 10 19 45.0 -00 16 34  & x  & 19.7  & 20.8  & Appenzeller treasure hunt ok \\
 42 & RXS J10244-0143  & 10 24 26.9 -01 43 01  & 0.093  &       & 3  & 62   & 23  & 10 24 28.2 -01 42 50  & x  & 15.9  & 17.0  & Appenzeller treasure hunt ok \\
 43 & RXS J10288+0139  & 10 28 48.6 +01 39 02  & 0.139  & 19.7  & 3  & 62   & 14  & 10 28 49.1 +01 38 50  & r  & 18.4  & 21.1  & Appenzeller treasure hunt ok \\
 44 & RXS J10359-0713  & 10 35 54.9 -07 13 07  & 0.223  & 19.7  & 3  & 62   & 37  & 10 35 55.8 -07 13 42  &    & 19.1  & 19.9  & Appenzeller treasure hunt ok \\
 45 & RXS J10397+0232  & 10 39 47.3 +02 32 02  & 0.155  & 19.8  & 3  & 62   & 29  & 10 39 47.9 +02 31 34  & x  & 18.0  & 20.2  & Appenzeller treasure hunt ok \\
 46 & RXS J10453-0346  & 10 45 23.7 -03 46 43  & 0.211  & 18.1  & 3  & 62   & 12  & 10 45 24.5 -03 46 46  & x  & 17.2  & 19.9  & Appenzeller treasure hunt ok \\
 47 & PC 1044+4719     & 10 47 13.2 +47 03 35  & 0.247  & 19.4  & 3  & 2043 & 9  & 10 47 13.2 +47 03 26  & r  & 17.8  & 19.5  & finding chart \\
 48 & RX J11125-1238   & 11 12 32.8 -12 38 30  & 0.142  & 17.6  & 3  & 120  & 16  & 11 12 32.6 -12 38 14  & x  & 16.3  & 17.5  & standout v ok, 1RXS J111233.1-123811 \\
 49 & IRAS11223-1244   & 11 24 50.0 -13 01 13  & 0.199  &       & 3  & 1214 & 11  & 11 24 50.7 -13 01 17  & r  & 16.6  & 18.5  & standout radio NVSS J112450.7-130118 \\
 50 & RXS J12212+3522  & 12 21 12.8 +35 22 54  & 0.299  & 18.9  & 3  & 62   & 13  & 12 21 13.6 +35 23 01  & x  & 19.1  & 19.9  & Appenzeller treasure hunt ok \\
 51 & IRAS 12202+1646  & 12 22 46.6 +16 29 43  & 0.181  & 18.0  & 3  & 1327 & 22  & 12 22 47.1 +16 29 30  &    & 16.8  & 18.3  &                         \\ 
 52 & Q 1246+4636      & 12 49 03.4 +46 20 30  & 0.089  & 21.4  & 3  & 2058 & 11  & 12 49 04.0 +46 20 39  &    & 20.0  & 20.4  & finding chart \\
 53 & PKS 1304-215     & 13 06 42.2 -21 48 12  & 0.127  & 17.5  & 3  & 258  & 21  & 13 06 42.0 -21 47 51  & r  & 16.1  & 18.4  & finding chart \\
 54 & IRAS 13305-1739  & 13 33 16.5 -17 55 00  & 0.148  & 17.2  & 3  & 1605 & 10  & 13 33 16.5 -17 55 10  & r  & 15.0  & 16.4  & finding chart \\
 55 & Q 1339.8+265     & 13 42 10.9 +26 13 10  & 0.148  & 20.1  & 3  & 512  & 9  & 13 42 10.3 +26 13 13  &    & 19.5  & 20.6  & obvious nearby \\
 56 & IRAS 13550-3142  & 13 57 55.3 -31 57 11  & 0.189  &       & 3  & 33   & 16  & 13 57 54.6 -31 56 58  & r  & 19.0  & 22.2  & radio NVSS J135754.6-315702 \\
 57 & WPVS 85          & 13 58 24.6 -38 09 17  & 0.034  &       & 3  & 2462 & 89  & 13 58 19.3 -38 10 21  &    & 18.2  & 18.6  & finding chart \\
 58 & KISSR 1751       & 14 16 45.2 +44 51 12  & 0.348  & 20.2  & 3  & 1549 & 7200 & 14 16 47.6 +42 51 13  &    & 18.2  & 20.8  & VCV off by 2 deg N/S \\
 59 & IRAS 14207-2002  & 14 23 33.2 -20 15 56  & 0.173  &       & 3  & 33   & 16  & 14 23 32.1 -20 15 51  & r  & 17.9  & 19.7  & radio nearby galaxy \\
 60 & IRAS 14317-3237  & 14 34 43.2 -32 50 25  & 0.025  & 14.4  & 3  & 575  & 30  & 14 34 45.5 -32 50 32  & rx  & 13.0  & 15.9  & finding chart \\
 61 & Q 14362-0657     & 14 38 53.0 -07 10 20  & 0.087  &       & 3  & 1609 & 24  & 14 38 54.4 -07 10 29  & x  & 16.9  & 18.2  & Einstein, only X-ray 1RXS J143854.0-071020 \\
 62 & MS 15325+0130    & 15 35 02.7 +01 20 58  & 0.074  & 18.0  & 3  & 1403 & 54  & 15 35 03.6 +01 20 06  & x  & 18.2  & 18.8  & finding chart \\
 63 & Q 1614+4635      & 16 15 50.9 +46 27 42  & 0.097  & 20.6  & 3  & 2058 & 8  & 16 15 51.4 +46 27 49  &    & 19.1  & 19.5  & finding chart \\
 64 & 1RXS J17082-0349 & 17 08 17.9 -03 49 16  & 0.180  & 17.5  & 3  & 1814 & 20 & 17 08 19.1 -03 49 25  &    & 20.2  & 21.6 & good fit to NVSS ellipse, 2007 A\&A-470-787, \\
    &                  &                       &        &       &    &      &    &    &    &    &    & OA confused RASS/NVSS/opt columns for 1RXS objects \\
 65 & RXS J17179+7308  & 17 17 58.0 +73 08 43  & 0.193  & 18.1  & 3  & 62   & 22  & 17 18 02.6 +73 08 52  & x  & 17.3  & 18.6  & Appenzeller treasure hunt ok \\
 66 & RXS J17325+7428  & 17 32 35.8 +74 28 08  & 0.207  & 19.6  & 3  & 62   & 27  & 17 32 39.6 +74 27 46  & x  & 18.8  & 20.2  & Appenzeller treasure hunt ok \\
 67 & RX J18020+6629   & 18 02 04.8 +66 29 13  & 0.265  &       & 3  & 825  & 14  & 18 02 07.0 +66 29 07  &    & 20.2  & 20.8  & only X-ray, FC 1996, MNRAS 281, 59 \\
 68 & VA-31032         & 18 54 01.3 -78 42 23  & 0.076  &       & 3  & 1067 & 162  & 18 54 57.3 -78 43 10  &    & 19.0  & 22.4  & approx, only radio galaxy in B1950 box \\
 69 & R021             & 19 13 00.0 -59 56 55  & 1.918  & 21.5  & 1  & 1896 & 495  & 19 12 23.2 -60 03 53  &    & 21.5  & 21.8  & finding chart Richer, H., 1978 ApJ L224, 9 \\ 
 70 & IRAS 19119-4952  & 19 15 45.2 -49 47 42  & 0.049  & 16.0  & 3  & 575  & 28  & 19 15 47.9 -49 47 34  & r  & 16.7  & 16.2  & finding chart \\
 71 & IRAS 19186-5748  & 19 22 50.7 -57 42 35  & 0.059  & 15.2  & 3  & 575  & 17  & 19 22 52.6 -57 42 28  &    & 15.0  & 16.3  & finding chart \\
 72 & 1ES 2055+298     & 20 58 01.2 +30 02 44  & 0.036  &       & 3  & 1775 & 183  & 20 58 12.3 +30 04 36  &    & 12.4  & 15.3  & Perlman treasure hunt ok, is IRAS 20561+2952 \\
 73 & OSMER 8          & 21 02 34.5 -39 49 48  & 0.246  & 18.5  & 3  & 1693 & 9  & 21 02 35.1 -39 49 43  &    & 17.0  & 18.5  & finding chart \\
 74 & RXS J21528+0523  & 21 52 50.0 +05 23 44  & 0.181  & 20.2  & 3  & 62   & 59  & 21 52 46.0 +05 23 53  &    & 20.0  & 20.7  & Appenzeller treasure hunt ok \\
 75 & 1E 22044+468     & 22 06 25.5 +47 04 43  & 0.163  & 19.7  & 3  & 1872 & 14  & 22 06 26.8 +47 04 48  &    & 18.8  & 19.9  & best nearby optical match \\
 76 & PKS 2206-251     & 22 09 24.2 -24 53 26  & 0.158  & 18.0  & 3  & 1949 & 25  & 22 09 22.9 -24 53 32  &    & 17.7  & 19.1  & OA states 2dF TGS061Z183 \\ 
 77 & RXS J22148+1351  & 22 14 52.9 +13 51 18  & 0.153  & 19.1  & 3  & 62   & 12  & 22 14 53.7 +13 51 19  &    & 18.8  & 19.5  & Appenzeller treasure hunt ok \\
 78 & IRAS 22152-0227  & 22 17 51.0 -02 22 29  & 0.093  & 17.0  & 3  & 2390 & 616  & 22 17 50.8 -02 12 25 & r  & 14.1 & 16.4  & VCV off 10 amin N/S, is NPM 1G-02.0483, NED ok \\
 79 & RXS J22367+0642  & 22 36 45.3 +06 42 16  & 0.269  & 20.0  & 3  & 62   & 11  & 22 36 45.2 +06 42 27  & x  & 19.6  & 20.4  & Appenzeller treasure hunt ok \\
 80 & RXS J22370+1235  & 22 37 00.8 +12 35 18  & 0.097  & 17.7  & 3  & 62   & 11  & 22 37 00.1 +12 35 12  & x  & 16.9  & 18.3  & Appenzeller treasure hunt ok \\
 81 & PKS 2242-29      & 22 44 51.8 -29 28 56  & 0.166  & 18.5  & 3  & 2010 & 11  & 22 44 52.6 -29 28 55  & r  & 17.0  & 19.9  & finding chart \\
 82 & IRAS 23515-2917  & 23 54 05.8 -29 01 00  & 0.335  & 19.7  & 3  & 33   & 10  & 23 54 06.5 -29 01 00  &    & 18.6  & 20.2  & nearest object, v ok \\
\hline
\end{tabular}}
\end{table*}

\section{Summary}

This paper presents 39 duplication removals, 380 astrometric moves of 8+ arcseconds, and 30 de-listings, to be applied to the V\'eron-Cetty \& V\'eron Quasar Catalogue, 13th edition.  This is to bring the VCV data up to the astrometric standard of today's large optical quasar surveys, and so enable accurate inclusion into dynamic databases like NED and SIMBAD (http://simbad.u-strasbg.fr/simbad).  

\section*{Acknowledgements}

Great thanks to Mira and Philippe V\'eron for their definitive catalogue, for critiques of my early work for this paper, and encouragement to publish.  Thanks to Adam Myers for XDQSO redshifts for some objects, and to Cedric Ledoux and Mike Irwin for help with single objects.  Thanks also to Steve Willner for supplying a copy of the Afanasiev paper.

\end{document}